\documentclass[12pt]{iopart} 
\usepackage{tikz}
\usepackage{graphicx}%
\usepackage{color} 
\usepackage{transparent}
\usepackage{psfrag}
\usepackage{dcolumn}
\usepackage{bm}
\usepackage{amssymb}
\usepackage{latexsym}
\usepackage{hyperref} 
\usepackage{booktabs}
\usepackage{latexsym}
\begin{document}
\newcommand{\tikzcircle}[2][red,fill=red]{\tikz[baseline=-0.5ex]\draw[#1,radius=#2] (0,0) circle ;}%
\def\bea{\begin{eqnarray}}
\def\eea{\end{eqnarray}}
\def\beq{\begin{equation}}
\def\eeq{\end{equation}}
\def\f{\frac}
\def\k{\kappa}
\def\e{\epsilon}
\def\ve{\varepsilon}
\def\be{\beta}
\def\D{\Delta}
\def\h{\theta}
\def\t{\tau}
\def\a{\alpha}

\def\cDa{{\cal D}[X]}
\def\cD{{\cal D}[x]}
\def\cL{{\cal L}}
\def\cLo{{\cal L}_0}
\def\cLa{{\cal L}_1}

\def\Re{{\rm Re}}
\def\sj{\sum_{j=1}^2}
\def\rk{\rho^{ (k) }}
\def\rek{\rho^{ (1) }}
\def\cek{C^{ (1) }}
\def\rz{\rho^{ (0) }}
\def\rt{\rho^{ (2) }}
\def\rtb{\bar \rho^{ (2) }}
\def\trk{\tilde\rho^{ (k) }}
\def\trek{\tilde\rho^{ (1) }}
\def\trz{\tilde\rho^{ (0) }}
\def\trt{\tilde\rho^{ (2) }}
\def\r{\rho}
\def\tD{\tilde {D}}

\def\rpl{r_\parallel}
\def\rp{{\bf r}_\perp}

\def\s{\sigma}
\def\kb{k_B}
\def\bF{\bar{\cal F}}
\def\F{{\cal F}}
\def\la{\langle}
\def\ra{\rangle}
\def\nn{\nonumber}
\def\up{\uparrow}
\def\dn{\downarrow}
\def\S{\Sigma}
\def\dg{\dagger}
\def\d{\delta}
\def\p{\partial}
\def\l{\lambda}
\def\L{\Lambda}
\def\G{\Gamma}
\def\o{\Omega}
\def\w{\omega}
\def\g{\gamma}

\def\bv{ {\bf b}}
\def\uv{ {\hat{\bm{u}}}}
\def\rv{ {\bf r}}
\def\vv{ {\bf v}}

\def\tbF{\tilde {\bm F}}
\def\tDv{\tilde {D_v}}
\def\tgv{\tilde {\gamma_v}}

\def\jv{ {\bf j}}
\def\jr{ {\bf j}_r}
\def\jd{ {\bf j}_d}
\def\jdd{ { j}_d}
\def\noi{\noindent}
\def\a{\alpha}
\def\d{\delta}
\def\p{\partial} 

\def\la{\langle}
\def\ra{\rangle}
\def\e{\epsilon}
\def\n{\eta}
\def\g{\gamma}
\def\break#1{\pagebreak \vspace*{#1}}
\def\hf{\frac{1}{2}}
\def\na{{\eta}_{\rm ac}}
\def\dac{D_{\rm ac}}
\def\n{{\eta}}
\def\gv{\gamma_v}


\title{Active Brownian motion with speed fluctuations in arbitrary dimensions: exact calculation of moments and dynamical crossovers
}

\author{Amir Shee}
\address{Institute of Physics, Sachivalaya Marg, Bhubaneswar 751005, India}
\address{Homi Bhaba National Institute, Anushaktigar, Mumbai 400094, India}
\ead{amir@iopb.res.in}

\author{Debasish Chaudhuri}
\address{Institute of Physics, Sachivalaya Marg, Bhubaneswar 751005, India}
\address{Homi Bhaba National Institute, Anushaktigar, Mumbai 400094, India}
\ead{debc@iopb.res.in}

\begin{abstract}
We consider the motion of an active Brownian particle with speed fluctuations in $d$-dimensions in the presence of both translational and orientational diffusion. We use an Ornstein-Uhlenbeck process for active speed generation. Using a Laplace transform approach, we describe and use a Fokker-Planck equation-based method to evaluate the exact time dependence of all relevant dynamical moments.  We present explicit calculations of such moments and compare our analytical predictions against numerical simulations to demonstrate and analyze several dynamical crossovers. The kurtosis of displacement shows positive or negative deviations from a Gaussian behavior at intermediate times depending on the dominance of speed or orientational fluctuations. 
\end{abstract}
\newpage
\tableofcontents

\maketitle

\section{Introduction}
Active matter consists of self-propelled units, each of which can consume and dissipate  internal or ambient energy to maintain the system out of equilibrium and generate systematic motion~\cite{Schweitzer2003, Marchetti2013, Bechinger2016, Romanczuk2012, Gompper2020}. The self-propulsion breaks the  detailed balance condition and the equilibrium fluctuation-dissipation relation. 
Examples of self-propelled entities abound in nature, ranging from motor proteins~\cite{Astumian2002, Reimann2002}, bacteria~\cite{Berg1972, Condat2005} to macro-scale entities like birds and animals~\cite{Niwa1994}. Inspired by natural examples, several artificial active elements have been fabricated. This includes colloidal microswimmers, active rollers, vibrated rods, and asymmetric disks~\cite{Marchetti2013, Bechinger2016}. Active colloids self-propel in their instantaneous heading direction through auto-catalytic drive utilizing ambient chemical, optical, thermal, or electric energy. They are typically modeled as active Brownian particles (ABP) with constant self-propulsion speed in a heading direction that undergoes orientational diffusion. Their long-time dynamics are similar to the run-and-tumble particles (RTP)~\cite{Cates2013} and the active Ornstein-Uhlenbeck process~\cite{Fodor2016,Das2018}. 
Despite enormous progress in the knowledge of collective properties of active matter, the non-equilibrium nature of individual particles are yet to be completely understood. 
Recent studies showed that even non-interacting self-propelled particles can display rich and counterintuitive physical properties~\cite{Sevilla2014, Grossmann2016, Kurzthaler2018, Basu2018, Malakar2018, Basu2019, Dhar2019, Shee2020, Santra2020a, Majumdar2020, Malakar2020, Basu2020, Basu2020a, Santra2021, Chaudhuri2021, Mori2021, Dean2021}. 
The RTP particles show a late-time condensation~\cite{Mori2021}. 
In the absence of thermal noise, exact short and long time properties of ABPs were obtained, and anisotropies in their short time motion were pointed out~\cite{Basu2018, Basu2019, Majumdar2020}. Such anisotropies survive even in the presence of thermal noise~\cite{Shee2020}.

In a collection of ABPs with constant self-propulsion, collisions can lead to  speed fluctuations~\cite{Redner2013, Fily2014} impacting their motion. In active polymers, the speeds of individual bond segments and the center of mass undergo  fluctuations due to bonding, bending, and self-avoidance~\cite{Isele-Holder2015, Winkler2020, Gupta2019, Shee2020}. Moreover, the generation of self-propulsion, be it via auto-catalysis in active colloids or complex active processes in motile cells, involves internal stochastic processes that render inherent fluctuations to active speed~\cite{Romanczuk2012, Schienbein1993, Schienbein1994, Schweitzer1998, Howse2007, Jiang2010, Pietzonka2018, Selmeczi2005, Frangipane2019, Otte2021}. The RTP model with generic speed distributions has been studied recently~\cite{Mori2021, DeBruyne2021, Mori2020a, Mori2020b, Gradenigo2019}.  Nevertheless,  apart from few exceptions~\cite{Schienbein1994, Peruani2007, Romanczuk2012}, in the most well studied ABP model, the active speed is taken to be constant.

In this paper, we reconsider the Schienbein-Gruler mechanism for active speed generation~\cite{Schienbein1993, Schienbein1994}. This involves an Ornstein-Uhlenbeck process leading to speed fluctuations around a well-defined mean.
The heading direction of self-propulsion undergoes orientational diffusion, as usual. In addition, we consider a translational thermal noise influencing the motion of ABPs. We extend and utilize a Fokker-Planck equation-based method~\cite{Shee2020, Hermans1952} to obtain arbitrary moments of the dynamics of speed-fluctuating ABPs in general $d$-dimensions. The presence of speed fluctuations and the competition between the speed relaxation time with other time scales, e.g., the persistence time, leads to new crossovers. The direct calculation presented here allows for coevolution of active speed generation and spatial displacement.  
Our general result for mean squared displacement when interpreted for two-dimensions  agrees with Ref.~\cite{Schienbein1993, Peruani2007} in the limit of fast relaxation of active speed, such that the steady-state limit of the speed correlation function can be used. 
We, in particular, analyze the changes in the physical properties of ABP due to speed fluctuations.
The analysis shows an intermediate time regime of clear sub-diffusive scaling in the positional fluctuations parallel to the initial heading orientation, a behavior that disappears in the limit of constant active speed.  The main achievements of this paper are the following: (i)~We discuss  a method for calculating the exact time-dependence of dynamical moments in arbitrary dimensions. We derive the expressions for the second and fourth moments of the displacement vector, its fluctuations, and fluctuations in its projection along the initial heading direction and in directions perpendicular to it. (ii)~We show and analyze the presence of multiple crossovers in the mean-squared displacement and fluctuations of displacement vectors. (iii)~In the intermediate time scales, the kurtosis of the displacement vector measuring the deviations from possible normal distributions changes between positive and negative values before returning to the Gaussian behavior at long times. Such deviations are controlled by the dominance of speed and orientational fluctuations, respectively.     

The paper is organized as follows. In Section~\ref{sec:model_abp_speed_fluct}, we describe the model. In Section~\ref{sec:Fokker_Planck_abp_speed_fluctuation}, we present  the Laplace transform method starting from the Fokker-Planck equation to derive the general  equation for calculating arbitrary moments of dynamical variables in $d$-dimensions. In the following Sections, we present calculations of particular quantities of interest: the mean speed and speed fluctuations in Section~\ref{sec:avg_speed_speed_fluct}, the speed, orientation, and velocity auto-correlation functions in Section~{\ref{sec:orientation_corr_speed_corr}}, and the mean-squared displacement and displacement fluctuations in Section~\ref{sec:disp_abp_speed_fluct}. In Section~\ref{sec:Fourth_moment_and_Kurtosis}, we calculate the fourth moment of displacement and the kurtosis to characterize the non-Gaussian nature of displacement fluctuations. The kurtosis shows positive and negative maxima in time corresponding to relaxations of speed and orientational fluctuations, respectively. Finally, in Section~\ref{sec:conclusions_ABP_speed_fluct}, we conclude by summarizing the main results.

\section{Model}
\label{sec:model_abp_speed_fluct}
An active Brownian particle (ABP) with fluctuating speed in $d$-dimension is described by its position $\rv = (r_1,r_2,\ldots,r_d)$ and active velocity $\bm{v}$ having a scalar speed $v$ in the  orientation $\uv=(u_1,u_2,\ldots,u_d)$, a $d$-dimensional unit vector that performs rotational diffusion on a unit sphere. The active speed $v$ is determined by a Ornstein-Uhlenbeck process. 
In the presence of a translational Brownian noise the motion of the particle is describes within the Ito convention~\cite{Ito1975, VandenBerg1985, Mijatovic2020} as, 
\bea
&dr_i = v(t) u_i \,dt +  dB_i^t(t),
\label{eom:disp_abp_speed_fluct}
\eea 
\bea
&dv = -\gv (v-v_0)\,dt + dB^s(t),
\label{eom:speed_abp_speed_fluct}
\eea 
\bea
&du_i = (\d_{ij}-u_i u_j)\,dB_j^r (t) - (d-1) D_r u_i \,dt.
\label{eom:rot_Ito_speed_fluct}
\eea 

Equation~(\ref{eom:disp_abp_speed_fluct}) describes the evolution of particle position due to time dependent active speed $v(t)$ in orientation $\uv(t)$. The stochastic variables $v(t)$ and $\uv(t)$ evolve independently. The translational diffusion due to thermal noise is described by  the Gaussian process $\bm{dB}^t$ with mean zero and variance $\la dB_i^t dB_j^t\ra = 2D\d_{ij}\, dt$. 

Equation~(\ref{eom:speed_abp_speed_fluct}) describes the active speed generation following an Ornstein-Uhlenbeck process~\cite{Schienbein1993, Schienbein1994}, with mean speed relaxing to $v_0$ in a time scale $\gv^{-1}$. The Gaussian stochastic process $dB^s$ obeys $\la dB^s(t) \ra = 0$ and $\la dB^s dB^s\ra = 2 D_v dt$, with $D_v$ governing the speed fluctuations. Note that this process does not always ascertain a positive speed. A large $D_v$ leads to larger fluctuations and as a result larger excursions towards negative speeds with respect to the heading direction. Here, it is instructive to note that such fluctuations of an effective negative speed can arise, e.g., in an assembly of repulsively interacting ABPs~\cite{Redner2013, Fily2014} due to increased frontal collisions at larger  particle density. In \ref{appendix:steady_state_prob_dist}, we discuss the dependence of cumulative speed distribution on the ratio $D_v/\gv$ for a given $v_0$.

Equation~(\ref{eom:rot_Ito_speed_fluct}) represents the orientational diffusion of the heading direction. The Gaussian white noise $\bm{dB}^r$ has  zero mean and variance $\la dB_i^r dB_j^r\ra = 2D_r\d_{ij}\, dt$. Alternatively, the equation can be expressed in the Stratonovich form $du_i = (\d_{ij}-u_i u_j) \circ dB_j^r (t)$.  Equation~(\ref{eom:rot_Ito_speed_fluct}) ensures the normalization ${\bf u}^2=1$ at all times.

\begin{figure*}[t]
\begin{center}
\includegraphics[width=12cm]{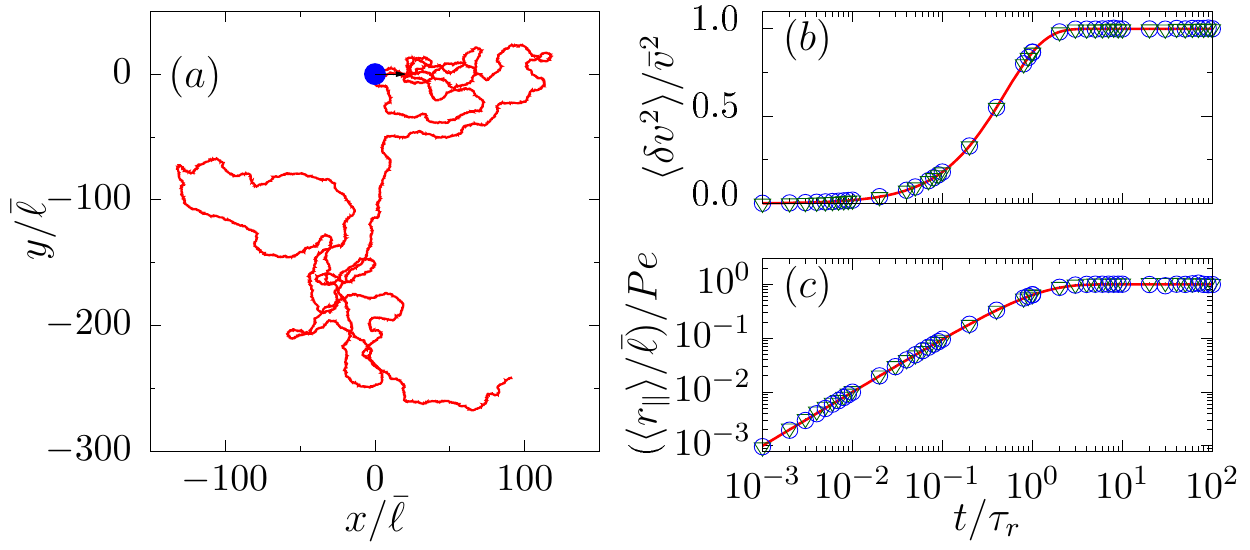} 
\caption{ (color online) ABP in two-dimensions (2d) with $\tDv=1$, and $\tgv=1$. $(a)$ Typical ABP trajectories over a duration $t=100\, \t_r$ for $Pe=20$. The blue point with arrow in each plot shows starting position and heading direction of the ABP. In these plots we used the initial active speed $v_1 = \bar v Pe$ and heading direction $\uv_0=\hat{x}$ along the $x$-axis. $(b)$ Scaled speed fluctuation $\la\d v^2\ra/\bar v^2$ as a function of time $t/\t_r$ for $Pe=1(\circ)$, $20(\triangledown)$. The points are simulation results and the solid line is a plot of equation~(\ref{eq:speed_fluctuation}). $(c)$ Displacement in the initial heading direction $\la r_{\parallel}\ra$ as a function of time $t$ for $Pe =1(\circ)$, $20(\triangledown)$. The points denote simulation results, and line depict $\la r_\parallel \ra = \la \rv \ra \cdot \uv_0$ using equation~(\ref{eq:ravg_speed_fluct}). } 
\label{fig:trajectory}
\end{center}
\end{figure*}

We set $\t_r = 1/D_r$ as the unit of time, and $\bar \ell = \sqrt{D/D_r}$ as the unit of length. All the speeds and velocities are expressed in units of $\bar v=\bar \ell/\t_r = \sqrt{D D_r}$. The dimensionless quantities controlling speed-fluctuation and speed-relaxation are $\tDv= D_v \t_r/\bar v^2 = D_v/D D_r^2$ and $\tgv = \g_v/D_r$. The mean active speed is expressed as a dimensionless Peclet number Pe\,$=v_0/\bar v = v_0/\sqrt{D D_r}$. 
It is straightforward to perform a direct numerical simulation of equations~(\ref{eom:disp_abp_speed_fluct}),~(\ref{eom:speed_abp_speed_fluct}), and~(\ref{eom:rot_Ito_speed_fluct}) using the Euler-Maruyama integration to generate trajectories as illustrated in figure~\ref{fig:trajectory}$(a)$.

\section{Calculation of moments from Fokker-Planck equation}
\label{sec:Fokker_Planck_abp_speed_fluctuation}
In this Section, we present a general framework for the calculation of arbitrary moments of dynamical variables~\cite{Shee2020, Chaudhuri2021}. 
The probability distribution $P(\rv, v, \uv, t)$ of the position $\rv$, the speed $v(t)$ and the heading direction $\uv$ of the particle  follows the Fokker-Planck equation
\bea
\fl
\p_t P(\rv, v, \uv, t) = D \nabla^2 P + D_r \nabla_u^2 P + D_v \partial_{v}^{2} P - v\, \uv\cdot \nabla P+\gv P 
+\gv (v-v_0) \partial_v  P
\eea
where $\nabla$ is the $d$-dimensional Laplacian operator, and $\nabla_u$ is the Laplacian in the ($d-1$) dimensional orientation space.
In terms of the Laplace transform $\tilde P(\rv, v, \uv, s) = \int_0^\infty dt\, e^{-s t}\, P(\rv, v, \uv, t) $, the Fokker-Planck equation takes the form,
\bea
\fl
-P(\rv, v, \uv, 0) + (s-\gv) \tilde P(\rv, v, \uv, s) &=& D \nabla^2 \tilde P + D_r \nabla_u^2 \tilde P +D_v\partial_{v}^{2} \tilde P 
- v\, \uv\cdot \nabla \tilde P+\gv (v-v_0)\partial_v \tilde P.  \nn
\eea
Defining the mean of an observable $\la \psi \ra_s = \int d\rv \,  dv\, d\uv\, \psi(\rv, v, \uv ) \tilde P(\rv, v, \uv, s)$, multiplying the above equation by $\psi(\rv, v, \uv)$ and integrating over all possible $(\rv, v, \uv)$  we obtain,
\bea
\fl
-\la \psi \ra_0 + s \la \psi \ra_s = D \la \nabla^2 \psi \ra_s + D_r \la \nabla_u^2 \psi \ra_s + D_v \la\p_{v}^{2}\psi\ra_s+ \la v\, \uv\cdot \nabla \psi \ra_s 
-\gv\la (v-v_0)\p_v \psi\ra_s. \nn\\
\label{eq:moment_recast}
\eea
where, the initial condition sets $\la \psi \ra_0 = \int d\rv \, dv\, d\uv\,  \psi(\rv, v, \uv) P(\rv, v, \uv, 0)$. 
Without any loss of generality, we consider the initial condition $P(\rv, v, \uv, 0) = \d(\rv) \d(v-v_1) \d(\uv - \uv_0)$, where $v_1$ is an initial speed that, in general, is different from $v_0$. 
Equation~(\ref{eq:moment_recast}) can be utilized to compute exact moments of any dynamical variable in $d$-dimensions as a function of time.
\section{Active speed}
\label{sec:avg_speed_speed_fluct}
In this Section, we first calculate the average active speed and speed fluctuations. We show how the speed fluctuations saturate over a long time. Next we calculate two-time auto-correlation functions for the heading direction, active speed, and velocity.  

\subsection{Mean speed}
To calculate the evolution of active speed, we use  $\psi=v$ and the initial condition $\la \psi \ra_0 = v_1$ in equation~(\ref{eq:moment_recast}). Other terms required for the calculation are: $\la\nabla^2 \psi\ra_s =0$, $\la\nabla_{u}^2 \psi\ra_s =0$, $\la\p_{v}^2 \psi\ra_s =0$, $\la v\uv\cdot\nabla \psi\ra_s =0$, $\la(v-v_0)\p_v \psi\ra_s = \la v\ra_s -v_0\la 1\ra_s=\la v\ra_s -v_0/s$. In the last relation we used $\la 1 \ra_s = \int d\rv~ d\uv ~dv \tilde P = \int d\rv ~d\uv ~dv \int_0^\infty dt~ e^{-st} P  = \int_0^\infty dt ~e^{-st} \{  d\rv~ d\uv~ dv P\} = \int_0^\infty dt~ e^{-st}  = 1/s$.  Thus from equation~(\ref{eq:moment_recast}), we get 
$\la v\ra_s= v_1/(s+\g_v) + v_0 \g_v/s(s+\g_v)$. 
The inverse Laplace transform of this relation gives 
\bea
\la v\ra(t)&=&v_1 e^{-\g_v t} + v_0 (1-e^{-\g_v t}). 
\label{eq_v}
\eea
At the long time limit of $\gv t \gg 1$ this gives the steady state value $\la v \ra = v_0$. 

\subsection{Speed fluctuations}

To calculate speed fluctuations, we consider $\psi=v^2$ and the initial condition $\la \psi \ra_0 = v_1^2$ in equation~(\ref{eq:moment_recast}).  The other terms involved in the calculation are: $\la\nabla^2 \psi\ra_s =0$, $\la\nabla_{u}^2 \psi\ra_s =0$, $\la\p_{v}^2 \psi\ra_s =\la 2 \ra_s = 2/s$, $\la v\uv\cdot\nabla \psi\ra_s =0$, $\la(v-v_0)\p_v \psi\ra_s = 2\la v^2\ra_s -2v_0\la v\ra_s$. Thus, we get from equation~(\ref{eq:moment_recast}), 
$\la v^2\ra_s = (s+2 \g_v)^{-1} [v_1^2 + 2 \g_v v_0 \la v \ra_s + 2D_v/s]$. 
Its inverse  Laplace transform gives 
\bea
\la v^2 \ra(t)&=& \left[ v_1 e^{-\g_v t} + v_0 \left( 1- e^{-\g_v t} \right)\right]^2 
+\f{D_v}{\g_v}  \left( 1- e^{-2 \g_v t} \right). 
\eea
As a result, using equation~(\ref{eq_v}),  the speed fluctuation can be expressed as, 
\bea
\la \d v^2\ra &=& \la v^2\ra-\la v\ra^2=\f{D_v}{\gv}\left(1-e^{-2\gv t}\right) .
\label{eq:speed_fluctuation}
\eea
This relation can be directly derived integrating equation~(\ref{eom:speed_abp_speed_fluct}) as is shown in  equation~(\ref{eq:speed_fluctuation_a}) of \ref{appendix:steady_state_corr_speed}. In the long time limit of $\gv t \gg 1$, the equation gives the steady state fluctuations $\la \d v^2\ra=D_v/\gv$. A comparison of the prediction of equation~(\ref{eq:speed_fluctuation}) with simulation results is shown in figure~\ref{fig:trajectory}$(b)$.

\section{Correlation functions}
\label{sec:orientation_corr_speed_corr}
The evolution of heading direction $\uv$ is an independent stochastic process, and thus does not get influenced by the speed fluctuations. Using equation~(\ref{eq:speed_fluctuation_a})  one can show that the persistence of heading direction decays as $\la \uv (t) \ra = \uv_0 e^{-(d-1)D_r t}$ and as a result the correlation
\bea
\la \uv (t) \cdot \uv (0) \ra &=& e^{-(d-1)D_r t}, 
\label{eq:orientational_corr_speed_fluct}
\eea
as was shown in Ref.~\cite{Shee2020}. The auto-correlation function of active speed can be directly calculated from equation~(\ref{eom:speed_abp_speed_fluct}) as is shown in \ref{appendix:steady_state_corr_speed},
\bea
\la \d v(t_1) \d v(t_2)\ra &=& \f{D_v}{\gv} \left[e^{-\gv | t_1 -t_2 |}-e^{-\gv(t_1+t_2)}\right],
\label{eq:autocorr_ABP_speed_fluct}
\eea    
where $\d v(t) = v(t) - \la v(t) \ra$. 
In the steady state limit of $t_1,\, t_2 \to \infty$, writing the time gap $\t= | t_1 -t_2 |$ one gets the simplified expression
\bea
\la \d v(\t) \d v(0)\ra &=& (D_v/\g_v) e^{-\g_v \t}.
\label{eq:speed_corr_function}
\eea 

The velocity correlation can be calculated directly from the Langevin equations writing $\dot \rv=\vv$. This gives $\la \vv(t) \ra = \la v(t) \ra \la \uv (t) \ra $ and $\la \vv(t_1) \cdot \vv(t_2) \ra = \la v(t_1) v(t_2)\ra \la \uv(t_1) \cdot \uv(t_2)\ra + 2 D \d(t_1-t_2)$. 
Thus, a direct calculation leads to 
\bea
\fl
\la\vv(t_1)\cdot \vv(t_2)\ra = \left[\f{D_v}{\gv} \left( e^{-\gv(t_1 -t_2)}-e^{-\gv(t_1 +t_2)}\right) + \la v (t_1) \ra \la v (t_2) \ra\right] e^{-(d-1)D_r (t_1 -t_2)} 
+ 2 D \d(t_1-t_2) \nn\\
\eea
The decay of velocity correlation is dictated by two time constants, the speed correlation time $\g_v^{-1}$ and the persistence time of the heading direction $D_r^{-1}$. 
The autocorrelation between fluctuations of velocity $\d \vv(t) = \vv(t) - \la \vv(t) \ra$ is given by $\la \d \vv(t_1) \d \vv (t_2) \ra$. 
Note that the mean velocity at time $t$ is given by $\la \vv (t)\ra = \la v (t) \ra \la \uv (t) \ra $, where $\la v (t) \ra$ is given by equation~(\ref{eq_v}) and $\la \uv (t) \ra = \uv_0 e^{-(d-1)D_r t}$. Therefore,
$
\la \vv (t)\ra = \left[ v_1 e^{-\g_v t} + v_0 (1-e^{-\g_v t})\right] \uv_0 e^{-(d-1)D_r t}. 
$
The expression simplifies in the steady state limit in which $\la v(t)\ra = v_0$, using $t_1,\,t_2 \to \infty$ and writing $t_1-t_2=\t$ one gets 
\bea
\la \d \vv(\t) \d \vv (0) \ra 
= \left[ v_0^2 + \f{D_v}{\g_v} e^{-\g_v \t}\right] e^{-(d-1)D_r \t} + 2 D \d(\t). 
\eea

\section{Displacement}
\label{sec:disp_abp_speed_fluct}
In this Section, we compute various moments of the displacement vector using equation~(\ref{eq:moment_recast}). 
We begin by setting $\psi = \rv$, and initial location $\la\psi\ra_0 = {\bf 0}$ at the origin. The calculation uses $\la \nabla^2 \psi \ra_s=0$, $\la \nabla_u^2 \psi \ra_s = 0$, $\la\p_{v}^{2}\psi\ra_s = 0$, $\la v\, \uv\cdot \nabla \psi \ra_s = \la v \uv \ra_s$, and $\la (v-v_0)\p_v \psi\ra_s = 0$ in equation~(\ref{eq:moment_recast}). This gives $\la \rv\ra_s=\la v \hat u\ra_s /s$. 
Again, the same equation~(\ref{eq:moment_recast}) gives $\la v \uv\ra_s= \f{1}{s+\gv+(d-1)D_r}\left[ v_1 \uv_0 + \gv v_0 \la\uv \ra_s \right]$ and $\la \uv \ra_s = \uv_0/(s+(d-1)D_r)$. Therefore, we get   
\bea
\la \rv \ra_s = \f{(v_1 -v_0) \uv_0}{s(s+(d-1)D_r+\gv)} +  \f{v_0 \, \uv_0}{s (s+(d-1)D_r)}. 
\label{eq_rvs}
\eea
Performing the inverse Laplace transform this leads to the evolution of the displacement vector 
\bea
\fl
\la \rv \ra (t) =   \f{(v_1-v_0) \uv_0}{(d-1)D_r+\gv} \left(1-e^{-(\, (d-1)D_r+\gv\,)\,t}\right) 
+ \f{v_0 \, \uv_0 }{(d-1)D_r}\left(1-e^{-(d-1)D_r t}\right) .
\label{eq:ravg_speed_fluct}
\eea
In Figure~\ref{fig:trajectory}$(c)$ we show a comparison of this estimate of displacement in the direction of the initial heading direction $\la r_\parallel \ra = \la \rv \ra \cdot \uv_0$ as obtained from equation~(\ref{eq:ravg_speed_fluct}) with numerical simulations.

{\bf Position-orientation cross-correlation $\la\uv\cdot\rv\ra$:}
Calculation of higher moments of displacement vector involves this equal time cross-correlation. 
We set $\psi = \uv\cdot\rv$ and the initial condition $\la\psi\ra_0 = 0$ in equation~(\ref{eq:moment_recast}). The calculation uses the relations:  
 $\la \nabla^2 \psi \ra_s=0$, $\la \nabla_u^2 \psi \ra_s = -(d-1)D_r\la\uv\cdot\rv\ra_s$, $\la\p_{v}^{2}\psi\ra_s = 0$, $\la v\, \uv\cdot \nabla \psi \ra_s = \la v \ra_s$, and $\la (v-v_0)\p_v \psi\ra_s = 0$.  As a result, one gets $\la \uv\cdot\rv\ra_s=\la v \ra_s/(s+(d-1)D_r)$. To completely determine the cross-correlation in the Laplace space,  we utilize equation~(\ref{eq:moment_recast}) further to obtain $\la v\ra_s= v_1/(s+\gv)+\gv v_0/s(s+\gv)$. These results lead to 
\bea
\la \uv\cdot\rv \ra_s = \f{v_1-v_0}{(s+\gv) (s+(d-1)D_r)}+\f{ v_0}{s(s+(d-1)D_r)}.
\label{eq:cross_corr_Laplace_abp_speed_fluct} 
\eea
The inverse Laplace transform of equation~(\ref{eq:cross_corr_Laplace_abp_speed_fluct}) gives 
\bea
\fl
\la \uv\cdot\rv \ra(t) = \f{v_1-v_0}{(d-1)D_r-\gv}\left( e^{-\gv t}-e^{-(d-1)D_r t}\right) 
+\f{v_0}{(d-1)D_r}\left(1-e^{-(d-1)D_r t}\right).
\label{eq:cross_corr_abp_speed_fluct}
\eea
It is interesting to note that  for initial active speed $v_1=v_0$, the cross-correlation reduces to $\la \uv\cdot\rv \ra(t) = v_0\left(1-e^{-(d-1)D_r t}\right)/(d-1)D_r$, an expression that is the same as ABPs in the absence of active speed fluctuations as described in Ref.~\cite{Shee2020}.
 
\subsection{Mean squared displacement}
\label{sec:msd_abp_speed_fluctuation}
Here we present an exact computation of the mean squared displacement $\la\rv^2\ra$. We use $\psi=\rv^2$ and the initial condition $\la \rv^2 \ra_0 = 0$ in equation~(\ref{eq:moment_recast}). The calculation of the moment uses the relations $\la \nabla_u^2 \rv^2 \ra_s =0$,  $\la \nabla^2 \rv^2 \ra_s = 2d/s$
and $\la v\,\uv\cdot \nabla \rv^2 \ra_s = 2 \la v\, \uv \cdot \rv \ra_s$. 
Thus equation~(\ref{eq:moment_recast}) leads to 
  \bea
  \la \rv^2 \ra_s =\f{1}{s} \left[ \f{2 d D}{s} + 2  \la v\, \uv \cdot \rv \ra_s\right] .
  \label{eq:r2_laplace}
  \eea  
  To complete the calculation, one needs to evaluate $\la v\, \uv \cdot \rv \ra_s $, again, using the same  equation~(\ref{eq:moment_recast}). One may proceed like before, using $\psi= v\, \uv \cdot \rv$ and $\la\psi\ra_0 = 0$, $\la\nabla^2\psi\ra_s = 0$, $\la\nabla_{u}^{2}(v\, \uv \cdot \rv)\ra_s = -(d-1)\la v \uv\cdot\rv\ra_s$, $\la v\uv \cdot \nabla (v\, \uv \cdot \rv) \ra_s = \la v^2 \uv^2 \ra_s=\la v^2\ra_s$ to obtain 
\bea  
   \la v\, \uv \cdot \rv \ra_s =\f{1}{(s+(d-1)D_r + \gv)} \left[\la v^2\ra_s+\gv v_0 \la \uv \cdot \rv \ra_s\right]. \nn
\eea  
   Similar calculations give,
   \bea
    \la v^2\ra_s=\f{1}{s+2\gv}\left[v_{1}^2 + 2D_v\la 1\ra_s +  \f{2\gv v_0}{s+\gv}\left[v_{1} + \gv v_0 \la 1\ra_s\right]\right],\nn\\
    \la \uv\cdot\rv \ra_s = \f{v_1-v_0}{(s+\gv) (s+(d-1)D_r)}+\f{ v_0}{s(s+(d-1)D_r)}. \nn
  \eea
  Thus, plugging these relations back in the expression of $\la \rv^2\ra_s$ in equation~(\ref{eq:r2_laplace}), we obtain
\bea
\la \rv^2\ra_s = \f{2 d D}{s^2} 
 + \f{4 D_v}{s^2 (s+2\gv)(s+(d-1)D_r+\gv)}\nn\\ +\f{2\gv v_0}{s(s+(d-1)D_r)(s+\gv)(s+(d-1)D_r+\gv)}\left(v_1 + \f{\gv v_0}{s}\right)\nn\\
 +\f{2}{s(s+(d-1)D_r+\gv)(s+2\gv)}\left[v_{1}^2 +\f{2\gv v_0}{(s+\gv)}\left(v_1 +\f{\gv v_0}{s}\right)\right]
 \label{eq:r2_Laplace_abp_speed_fluct}
\eea
Performing the inverse Laplace transform, this leads to
\bea
\la \rv^2 \ra = \frac{\left(D_{v}-\gv (v_0-v_1)^2\right)e^{-2
\gv t} }{\gv^2 ((d-1) D_{r}-\gv)}+\frac{2  (2
(d-1) D_{r}-\gv) v_0 (v_0-v_1) e^{-\gv t}}{(d-1) D_{r} \gv ((d-1) D_{r} - \gv)}\nn\\
 +\frac{2  \left(-\gv^2 v_0^2 +(d-1) D_{r} \gv v_0 v_1\right)e^{-(d-1)
D_{r} t}}{(d-1)^2 D_{r}^2 ((d-1) D_{r}-\gv)
\gv}\nn\\
+\frac{2
\left((d-1)^2 d D D_{r}^2 \gv+\gv^2 v_0^2+(d-1) D_{r} \left(D_{v}+\gv
\left(d D \gv+v_0^2\right)\right)\right)t}{(d-1) D_{r} \gv ((d-1) D_{r}+\gv)}\nn\\+\frac{2
(d-1) D_{r} \gv^3 v_0 (-3 v_0+v_1)-2 \gv^4 v_0^2}{(d-1)^2 D_{r}^2 \gv^2
((d-1) D_{r}+\gv)^2}\nn\\
-\frac{(d-1)^3 D_{r}^3 (D_{v}+\gv (v_0-v_1) (3 v_0+v_1))+(d-1)^2
D_{r}^2 \gv \left(3 D_{v}+\gv \left(7 v_0^2-4 v_0 v_1-v_1^2\right)\right)}{(d-1)^2 D_{r}^2 \gv^2
((d-1) D_{r}+\gv)^2}\nn\\
-\frac{2 \left((d-1) D_{r} \gv (2 D_{v}-\gv (v_0-v_1) (v_0-v_1))\right)e^{-((d-1) D_{r}+\gv) t} }{(d-1) D_{r} ((d-1) D_{r}-\gv) \gv ((d-1) D_{r}+\gv)^2}\nn\\
 -\frac{2 \left((d-1)^2 D_{r}^2 \gv (v_0-v_1)
v_1+\gv^3 v_0 (-v_0+v_1)\right)e^{-((d-1) D_{r}+\gv) t} }{(d-1) D_{r} ((d-1) D_{r}-\gv) \gv ((d-1) D_{r}+\gv)^2}
\label{eq:r2_abp_speed_fluct_v1}
\eea
The derivation of $\la \rv^2\ra$ in $d$-dimensions shown in equation~(\ref{eq:r2_abp_speed_fluct_v1}) is our first main result.
Considering the initial active speed $v_1=v_0$,  equation~(\ref{eq:r2_abp_speed_fluct_v1}) simplifies to
\bea
\la \rv^2 \ra &=& 2 d D t + \f{2v_0^2}{(d-1) D_r} \left( t  - \f{1-e^{-(d-1)D_r t}}{(d-1) D_r}\right) \nn\\
&+&\f{2 D_v}{\gv(\gv+(d-1)D_r)}\left( t - \f{1 -  e^{-(\gv+(d-1)D_r)t}}{\gv+(d-1)D_r} \right) \nn\\
&-& \f{2 D_v}{\gv(\gv+(d-1)D_r)}\left[  \f{1-e^{-2\gv t}}{2\gv} - \f{e^{-2\gv t} - e^{-(\gv+(d-1)D_r) t}}{\g_v-(d-1)D_r}\right] .
\label{eq:r2_abp_speed_fluct}
\eea
Note that for the special case of $(d-1)D_r = \gv$, equation~(\ref{eq:r2_abp_speed_fluct}) can be further simplified by using the L'H{\^o}pital's rule, or, directly substituting $(d-1)D_r = \gv$ and $v_1=v_0$  in equation~(\ref{eq:r2_Laplace_abp_speed_fluct}) to calculate $\la \rv^2 \ra$.

In the limits of $D_v \to 0$ and $\g_v \to \infty$, equation~(\ref{eq:r2_abp_speed_fluct}) reduces to that of free ABPs in the absence of speed fluctuations, as shown in Ref.~\cite{Shee2020}. 
The structure of the second and third terms in equation~(\ref{eq:r2_abp_speed_fluct}) can describe two ballistic diffusive crossovers~\cite{Peruani2007}. As we show in the following, the presence of the fourth term allows for further crossovers. Moreover, the presence of translational diffusion makes the short time dynamics diffusive.
Here, it is instructive to note that the calculations of lower moments can be performed easily using the Langevin equations. For example, the formal solution for the position vector,  $\rv(t) = \int_{0}^{t} dt^{\prime} v(t^{\prime}) \uv(t^{\prime})+\int_{0}^{t} \bm{dB}^t(t^{\prime})$ leads to the second moment   
\bea
\fl
 \la \rv^2 \ra = \int_{0}^{t} dt_1 \int_{0}^t dt_2 \, \la v(t_1) v(t_2)\ra \, \la \uv(t_1)\cdot\uv(t_2)\ra 
 + \int_{0}^{t} \int_{0}^t \la \bm{dB}^t(t_1) \cdot \bm{dB}^t(t_2)\ra, 
 \label{eq:r2d2_integration_persistent_speed_fluct}
 \eea
where the cross terms do not appear as they describe independent stochastic processes with $\la \bm{dB}^t\ra=0$. By substituting the speed correlation function from equation~(\ref{eq:autocorr_ABP_speed_fluct}) and the orientational correlation function from  equation~(\ref{eq:orientational_corr_speed_fluct}) in equation~(\ref{eq:r2d2_integration_persistent_speed_fluct}), and performing the integrations, one gets  the same mean squared displacement relation as in equation~(\ref{eq:r2_abp_speed_fluct}). 

\begin{figure*}[t]
\begin{center}
\includegraphics[width=12cm]{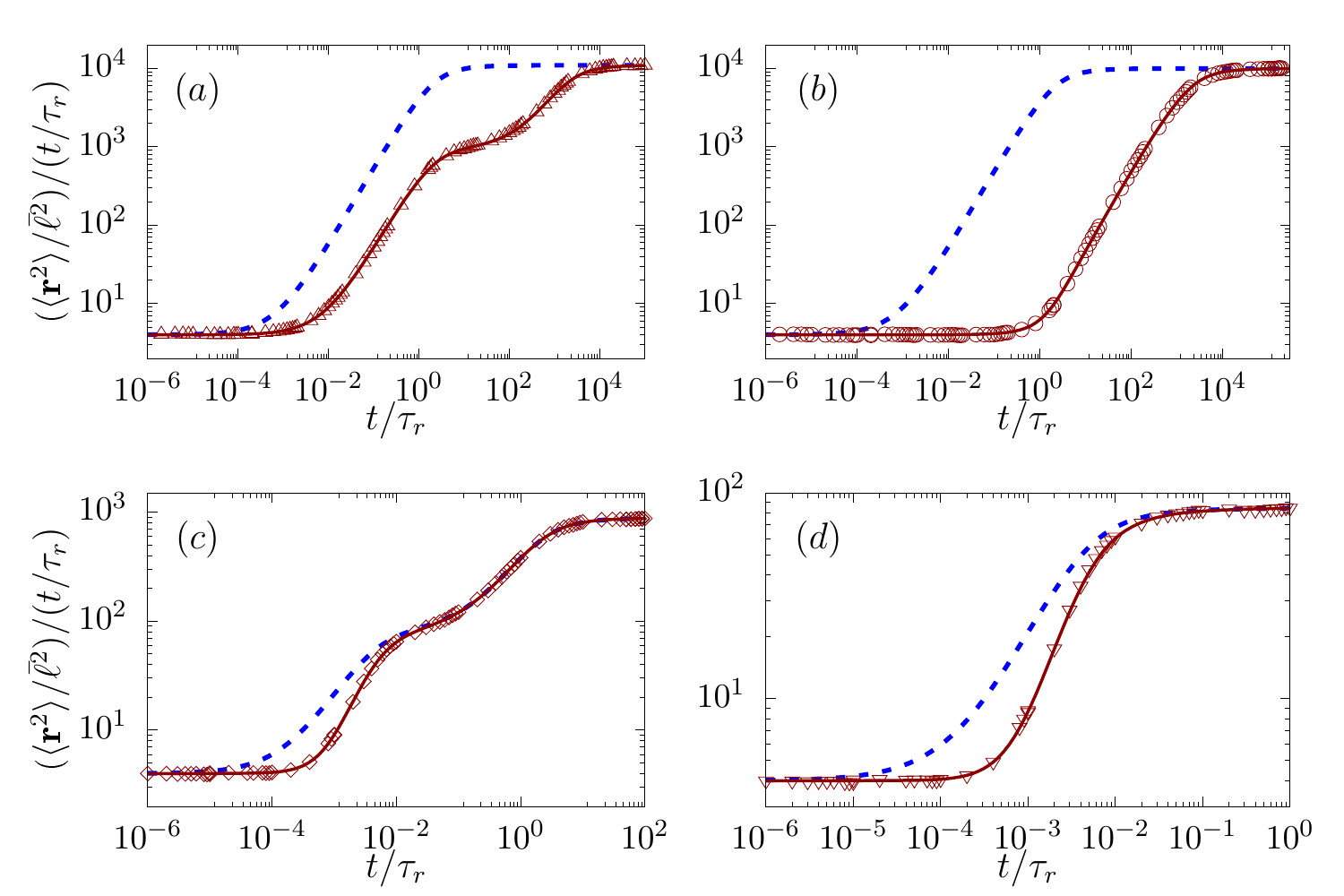} 
\caption{ (color online) Time dependence of $\la \rv^2\ra/t$ in 2d. The slow and fast relaxations of active speed are considered in $(a,b)$  $\tgv\ll1$, and $(c,d)$ $\tgv\gg1$, respectively. The points denote simulation results, the solid lines depict equation~(\ref{eq:r2_abp_speed_fluct}) with $d=2$, and the dashed lines depict equation~(\ref{eq:r2d2_abp_speed_fluct_old}). 
Parameter values used in $(a,b)$: $\tgv=5\times 10^{-4}$, $\tDv=2.5$ with $Pe = 22.36$ 
$(a)$ and $1.12$ $(b)$. 
Parameter values used in $(c,d)$: $\tgv=5\times 10^{2}$, $\tDv=10^{7}$ with $Pe = 20$ $(c)$ and $1$ $(d)$. Initial conditions are chosen such as the active speed speed $v_1/\bar{v}=Pe$ and the heading direction $\uv_0=\hat{x}$ is along the $x$-axis. 
} 
\label{fig:r2avg_comparision}
\end{center}
\end{figure*}

In figure\ref{fig:r2avg_comparision}  we compare our analytic prediction for the second moment of displacement shown in equation~(\ref{eq:r2_abp_speed_fluct}) with direct numerical simulation results in 2d ($d=2$) to find excellent agreement between them. Here, it is instructive to note the difference of our $d$-dimensional expression for $\la \rv^2\ra$ shown in equation~(\ref{eq:r2_abp_speed_fluct}) from earlier results for 2d obtained in Ref.~\cite{Schienbein1993, Peruani2007}. The difference stems from an assumption of time-scale separation used in these earlier publications, where the speed fluctuations were assumed to be in steady state. This can be easily seen by noting that 
instead of using the general result for $\la v(t_1) v(t_2)\ra$ of equation~(\ref{eq:autocorr_ABP_speed_fluct}), if one uses the steady state limit of the correlation for active speed as in equation~(\ref{eq:speed_corr_function}), the expression in equation~(\ref{eq:r2d2_integration_persistent_speed_fluct}) leads to the previously obtained relation for the second moment of displacement~\cite{Schienbein1993, Peruani2007}
\bea
\fl
\la \rv^2 \ra = 4D\,t+2 v_0^2\left(\f{t}{D_r} -\f{1-e^{-D_r t}}{D_r^2} \right) 
+\f{2 D_v}{\gv} \left( \f{t}{(\gv+D_r)} -\f{1-e^{-(\gv+D_r)t}}{(\gv+D_r)^2}\right).
\label{eq:r2d2_abp_speed_fluct_old}
\eea
As is clearly shown in figure\ref{fig:r2avg_comparision}, while our calculation in  equation~(\ref{eq:r2_abp_speed_fluct}) exactly captures the behavior observed in numerical simulations, the earlier result shown in equation~(\ref{eq:r2d2_abp_speed_fluct_old}) deviates from the numerically obtained $\la \rv^2\ra(t)$. In figure~\ref{fig:r2avg_comparision},  the qualitative difference can be seen clearly at small $\g_v/D_r$ and large $Pe$.  The figure  shows multiple ballistic diffusive crossovers, which we describe in detail in the following. 

{\bf Multiple crossovers and crossover timescales:}
To elucidate the crossovers permitted by equation~(\ref{eq:r2_abp_speed_fluct}), we focus on its behavior in different time regimes. First, we note that in the two limits of the shortest and longest times $\la \rv^2\ra$ shows diffusive behavior, albeit with two significantly different diffusion constants. In the short time limit
\bea
\la \rv^2 \ra \approx 2d \, D\, t  
\label{eq:diff1}
\eea 
and in the long time limit
\bea
\la \rv^2 \ra \approx 2d \left[D + \f{v_0^2}{d(d-1)D_r} + \f{D_v}{d \g_v [\g_v+(d-1)D_r]} \right] t.
\label{eq:diff2}
\eea
For the smallest time scales, we expand $\la \rv^2\ra$ in equation~(\ref{eq:r2_abp_speed_fluct}) around $t=0$ to obtain  
\bea
\la \rv^2 \ra &=& 2dDt + v_0^2 t^2 -\f{1}{3} \left((d-1)D_r v_0^2 -2 D_v\right)t^3+{\cal O}(t^4).
\label{eq:r2_expansion_small_time}
\eea
This shows a crossover from diffusive $\la \rv^2\ra \sim t$ to ballistic behavior $\la \rv^2\ra \sim t^2$ at $t_{I} = 2dD/v_0^2$, with the crossover point obtained by comparing the first and second terms of the above expansion. Such crossovers have been observed in figure~\ref{fig:r2avg_comparision}. Comparing the second and third terms in the above expansion, one can identify a possible second crossover from ballistic to diffusive behavior at  $t_{II}= 3 v_0^2 / [ (d-1)D_r v_0^2-2D_v ]$, provided $(d-1)D_r v_0^2 > 2 D_v$. 
Further insights can be drawn by separately considering the limits of (i)~slow speed relaxation and (ii)~slow orientational relaxation, separately. 

(i)~{\bf Slow  relaxation of active speed; $\g_v \ll (d-1)D_r$ :} Using  $(d-1)D_r t \gg 1$ and $2\gv t \ll 1$, we can write $\exp[{-(d-1)D_r t}] \approx 0$, $\exp[{-((d-1)D_r+\gv)t}] \approx 0$ and expand $\exp({-2\gv t})$ around $2\gv t= 0$ in equation~(\ref{eq:r2_abp_speed_fluct}) to get 
\bea
\fl
\la \rv^2 \ra = \left(2dD+ \f{2 v_0^2}{(d-1)D_r}+ \f{4D_v}{(d-1)^2 D_r^2 -\gv^2}\right) t 
+ \f{2D_v}{(d-1)D_r-\gv} t^2 +{\cal O}(t^3). 
\label{eq:r2_expan_int_dr_gt_gv}
\eea
This implies a possible third crossover $\la\rv^2\ra\sim t$ to $\sim t^2$ expected at $t_{III} \sim [2dD + 2v_0^2/(d-1)D_r +4D_v/((d-1)^2 D_r^2-\gv^2)]\,  [(d-1)D_r-\gv]/2D_v$. 
The final crossover point to the long-time diffusive limit denoted by equation~(\ref{eq:diff2}) can be calculated by comparing the last term in equation~(\ref{eq:r2_expan_int_dr_gt_gv}) with equation~(\ref{eq:diff2}). This crossover time turns out to be $t_{IV} \sim [2dD+ 2 v_0^2/(d-1)D_r + 2D_v/\gv((d-1)D_r+\gv)] \, [(d-1)D_r - \gv)]/2D_v$.

Moreover, at small $Pe$, the diffusive-ballistic crossover at $t_I$ can be preempted by a different ballistic-diffusive crossover at $t^\ast_I$ that can be determined by comparing the first term in equation~(\ref{eq:r2_expansion_small_time}) with the second term in equation~(\ref{eq:r2_expan_int_dr_gt_gv}). This gives $t^\ast_I = d D [(d-1)D_r - \g_v]/D_v$, a crossover point independent of the active speed $v_0$.  

{Such crossovers for $\la\rv^2\ra$ in 2d, in the limit of $\gv\t_r \ll 1$, are illustrated in figure~\ref{fig:r2avg_abp_speed_fluct}.$(a)$. The graphs depict the expression in equation~(\ref{eq:r2_abp_speed_fluct}) using parameter values $\tgv = \gv\t_r=5\times 10^{-4}$, $\tDv = D_v \t_r/\bar{v}^2=2.5$. The solid line at larger $Pe\, (\,= 22.36)$ shows all four diffusive- ballistic- diffusive crossovers discussed above, as the requirement $t_{I}<t_{II}<t_{III}<t_{IV}$ is satisfied. In this case, the crossover times are 
$t_{I}/\t_r\sim 4/Pe^2\approx 0.008$, 
$t_{II}/\t_{r}
= 3 Pe^2/ (Pe^2- 2 \tDv) 
\approx 3.03$, 
$t_{III}/\t_r
= [4 + 2Pe^2+4 \tDv /(1-\tgv^2) ] (1- \tgv) / 2 \tDv 
\approx 202.8$, 
and 
$t_{IV}/\t_{r}
= [4+ 2 Pe^2 + 2 \tDv/\{ \tgv (1+ \tgv)\} ]\, (1 -  \tgv)/2 \tDv
\approx 2200$, 
as pointed out in figure~\ref{fig:r2avg_abp_speed_fluct}$(a)$. 

{For $Pe = v_{0}/\bar{v}= 1.12$, $\la \rv^2\ra$ 
denoted by the dashed line in  figure~\ref{fig:r2avg_abp_speed_fluct}.$(a)$ shows only two crossovers: ($a$)~a diffusive-ballistic crossover at 
$t^\ast_I/\t_r = 2(1-\tgv)/\tDv =0.8$ 
and 
($b$)~a ballistic-diffusive crossover at $t_{IV}/\t_r \approx 2000$. In this case $t^\ast_I < t_I=3.2 \t_r$, thus the first diffusive-ballistic crossover is preempted by $t^\ast_I$. 
Other possible intermediate crossovers disappear due to the following reasons. The possible ballistic-diffusive crossover point $t_{II} <0$ for these parameters. In its absence, the point $t_{III} \approx 3.3 \t_r$ signifying a possible diffusive-ballistic crossover cannot show any change in the already ballistic property of the ABP in that time regime. }   

\begin{figure*}[t]
\begin{center}
\includegraphics[width=8cm]{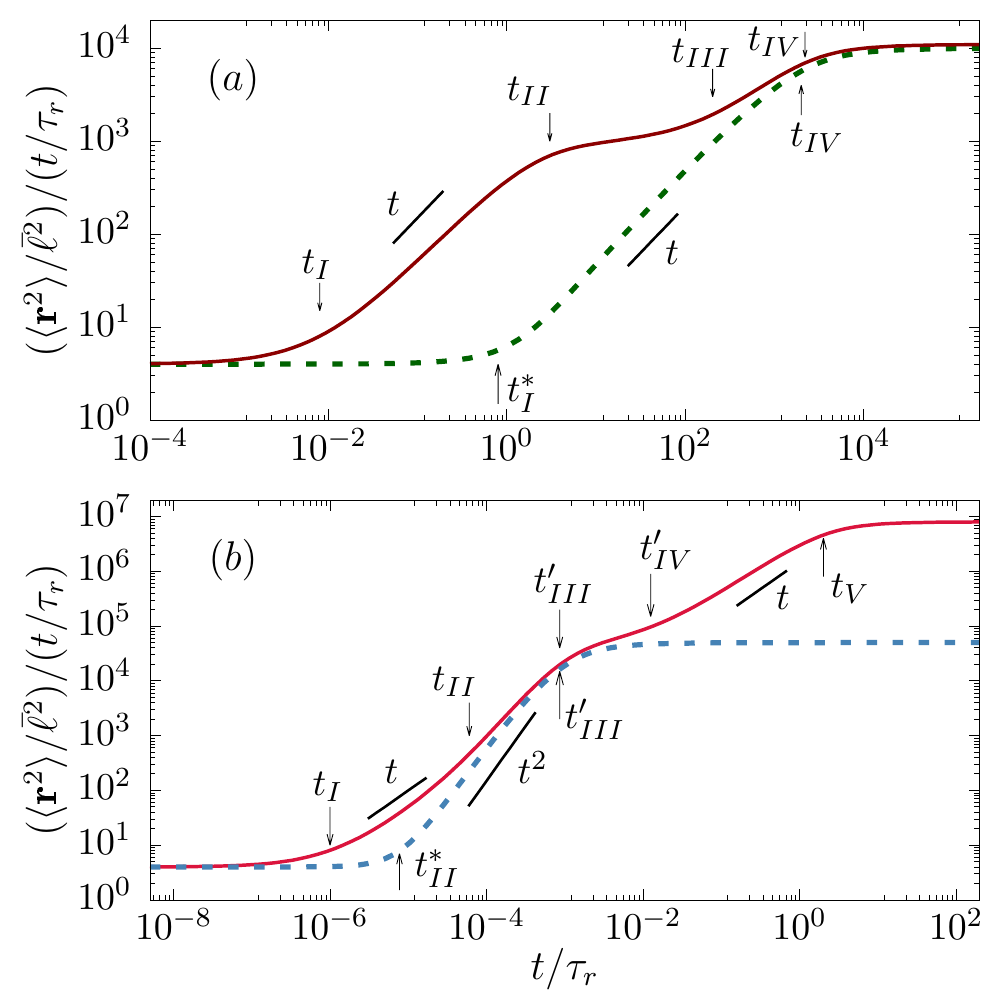} 
\caption{ (color online)  Persistent motion. Mean squared displacement $\la \rv^2\ra$ as in equation~(\ref{eq:r2_abp_speed_fluct}) as a function of time $t$ in two dimension, $d=2$. $(a)$ Parameters used are $\tgv=5\times 10^{-4}$, $\tDv=2.5$ with $Pe =22.36$~(solid line), $1.12$~(dashed line). The solid line shows four crossover with crossover times $t_{I}/\t_r=0.008$, $t_{II}/\t_r=3.03$, $t_{III}/\t_r=202.8$ and $t_{IV}/\t_r=2200$. The dashed line shows two crossovers: : a diffusive-ballistic crossover at $t^\ast_{I}/\t_r=0.8$ and a ballistic-diffusive crossover at $t_{IV}/\t_r=2000$. 
$(b)$ Parameters used are $\tgv=2\times 10^{3}$, $\tDv=10^{11}$,  with $Pe=2\times 10^{3}$~(solid line), $10$~(dashed line). The solid line shows five crossover with crossover times $t_{I}/\t_r=10^{-6}$, $t_{II}/\t_r=6 \times 10^{-5}$, $t^\prime_{III}/\t_r=8.6 \times 10^{-4}$ and $t^\prime_{IV}/\t_r=1.25 \times 10^{-2}$, and $t_V=2.02$. The dashed line shows two crossovers with crossover times 
$t^\ast_{II}/\t_r=7.75\times 10^{-6}$ and $t^\prime_{III}/\t_r=8.6\times 10^{-4}$. 
Initial activity: speed $v_1/\bar{v}=Pe$ and heading direction along $x$-axis, $\uv_0=\hat{x}$. 
} 
\label{fig:r2avg_abp_speed_fluct}
\end{center}
\end{figure*}

(ii)~{\bf Fast relaxation of active speed; $\gv \gg (d-1)D_r$}: The scenario of short-time diffusive-ballistic crossover at $t_{I}=2dD/v_0^2$ remains unchanged. As indicated before,  at $t_{II}= 3 v_0^2 / ( (d-1)D_r v_0^2-2D_v)$ with $2D_v>(d-1)D_r v_0^2$, a possible second crossover from $\la\rv^2\ra\sim t^2$ to $\la\rv^2\ra\sim t^3$ can appear. In the limit of $(d-1)D_r t\ll1$ and $2\gv t\gg1$, we can use $\exp({-2\gv t}) \approx 0$, $\exp[{-((d-1)D_r+\gv)t}] \approx 0$ and expand $\exp[-(d-1) D_r t]$ around $D_r t= 0$ in equation~(\ref{eq:r2_abp_speed_fluct}) to get 
\bea
\la \rv^2 \ra = \left(2dD+\f{2D_v}{\g_v((d-1)D_r +\gv)}\right) t + v_0^2 t^2 +{\cal O}(t^3).
\label{eq:r2_expan_int_dr_lt_gv}
\eea

Comparing the third term in equation~(\ref{eq:r2_expansion_small_time}) and the first term in  equation~(\ref{eq:r2_expan_int_dr_lt_gv}), we estimate the crossover time from $\la\rv^2\ra\sim t^3$ to $\la\rv^2\ra\sim t$ to be at $t^\prime_{III} = \left[3\f{2dD+2D_v/ \{\g_v((d-1)D_r +\gv)\}}{2D_v-(d-1)D_r v_0^2}\right]^{1/2}$. Equation~(\ref{eq:r2_expan_int_dr_lt_gv}) suggests a fourth crossover from $\la\rv^2\ra\sim t$ to $\la\rv^2\ra \sim t^2$ at $t^\prime_{IV} = (2dD +2D_v/(\g_v((d-1)D_r +\gv))/v_0^2$. The final ballistic-diffusive crossover with $(d-1)D_r\ll 2\gv$ can be obtained by comparing the second  term in equation~(\ref{eq:r2_expan_int_dr_lt_gv}) with equation~(\ref{eq:diff2}). This gives the final crossover time $t_{V} = (2dD+ 2 v_0^2/(d-1)D_r + 2D_v/\gv((d-1)D_r+\gv))/v_0^2$.

There is also a possibility of getting a direct crossover from $\la \rv^2 \ra \sim t$ to $\la \rv^2 \ra \sim t^3$ at $t^\ast_{II} = \left[6dD/(2D_v-(d-1)D_r v^2_{0})\right]^{1/2}$ if $t^\ast_{II} < t_{I}$. The crossover point $t^\ast_{II}$ is obtained by comparing the first and the third term of equation~(\ref{eq:r2_expansion_small_time}). Another direct final crossover from  $\la \rv^2 \ra \sim t^3$ to  $\la \rv^2 \ra \sim t$ can appear at $t_{VI}= \{6d[D + \f{v_0^2}{d(d-1)D_r} + \f{D_v}{d \g_v [\g_v+(d-1)D_r]}]/[2D_v - (d-1)D_r v_0^2]\}^{1/2}$ if $t_{VI} < t^\prime_{III}$, otherwise the final crossover will be at $t^\prime_{III}$. The estimate of  $t_{VI}$ is obtained by comparing the third term in equation~(\ref{eq:r2_expansion_small_time}) with equation~(\ref{eq:diff2}).

Such crossovers for $\la\rv^2\ra$ in 2d in the limit of $\gv\t_r \gg 1$ are illustrated in figure~\ref{fig:r2avg_abp_speed_fluct}.$(b)$. The graphs depict the expression in equation~(\ref{eq:r2_abp_speed_fluct}) using parameter values $\tgv = \gv\t_r=2\times 10^3$, $\tDv=D_v\t_r/\bar{v}^2=10^{11}$. The solid line at $Pe = v_{0}/\bar{v}=2\times 10^3$ 
exhibits all  five possible crossovers from $\la\rv^2\ra\sim t$ to $\sim t^2$, to $\sim t^3$, to $\sim t$, to $\sim t^2$ to finally $\sim t$ as the requirement $t_{I}<t_{II}<t^\prime_{III}<t^\prime_{IV}<t_{V}$ is satisfied. The crossover times are 
$t_{I}/\t_r\sim 4/Pe^2 =1\times 10^{-6}$, 
$t_{II}/\t_r \sim 3 Pe^2 / (Pe^2-2 \tDv^2)\approx 6\times 10^{-5}$, 
$t^\prime_{III}/\t_r = \left[3\left(4+2 \tDv/ \{ \tgv(1 + \tgv) \}\right)/(2 \tDv -Pe^2)\right]^{1/2} 
\approx 8.6\times 10^{-4},$ 
 $t^\prime_{IV}/\t_r
= [4+2 \tDv/ \{\tgv(1+ \tgv)\}]/Pe^{2} 
\approx 1\times 10^{-2}$, 
$t_{V}/\t_r
= [4+ 2 Pe^2 + 2 \tDv/ \{\tgv(1+ \tgv)\}]/Pe^2 \approx 2.$ 
They are identified by arrows on the solid line in figure~\ref{fig:r2avg_abp_speed_fluct}$(b)$. 

{For $Pe = v_{0}/\bar{v}=10$, $\la\rv^2\ra$ denoted by  the dashed line in  figure~\ref{fig:r2avg_abp_speed_fluct}$(b)$ shows only two crossovers: the first from $\la\rv^2\ra\sim t$ to $\la\rv^2\ra\sim t^3$ at $t^\ast_{II}$, and the second going back to $\la\rv^2\ra \sim t$ at $t_{VI}$. Here,  
$t^\ast_{II}/\t_{r}
= [12/( 2\tDv - Pe^2 ) \,]^{1/2}
\approx 7.75\times 10^{-6}$ as $t_{II}^{*}<t_{I}\approx 0.04\, \t_r$. 
The $\la \rv^2\ra \sim t^3$ to $\la \rv^2\ra \sim t$ crossover appears at  $t_{III}^{\prime}/\t_r \approx 8.6\times 10^{-4}$, as $t_{VI}/\t_r = \left[12 \times \f{1+\f{Pe^2}{2} + \f{\tDv}{\tgv(1+\tgv)}}{2 \tDv - Pe^2} \right]^{1/2} = 10^{-3} > t^\prime_{III} $. }
We list the dominance of different kinds of fluctuations in different time regimes in table~\ref{table1}.

\begin{table}[h]
\caption{$\la\rv^2\ra$ scaling: characterizing dominance of fluctuation in diffrent regime}
\centering
Direction of increasing time $t\longrightarrow$\\~\\
\begin{tabular}{| l | l | l | l | l | l |}
\hline
~~~~~~~~~~~ $\la\rv^2\ra$ & $\sim t$ & $\sim t^2$ & $\sim t$ & $\sim t^2$ & $\sim t$\\
\hline
 &  &  & thermal + &  & thermal + \\
$(d-1)D_r \gg\gv$ & thermal & orientation & orientation & speed &  orientation + \\
 &  &  &  &  & speed \\
\hline
 &  &  & thermal + &  & thermal + \\
$(d-1)D_r \ll\gv$ & thermal & speed & speed & orientation &  speed + \\
 &  &  &  &  & orientation \\
\hline
\end{tabular}
\label{table1}
\end{table}%


\subsection{Displacement fluctuations}
\label{sec:disp_fluct_abp_speed_fluct}
In this Section, we compute the displacement fluctuation $\la\d\rv^2\ra$ and analyze the multiple crossovers that it shows identifying the crossover times.
The displacement fluctuation is defined as $\la \d \rv^2\ra = \la \rv^2\ra - \la \rv \ra^2$ where $\la\rv^2\ra$ and $\la\rv\ra$ were already calculated in equations~(\ref{eq:r2_abp_speed_fluct}) and (\ref{eq:ravg_speed_fluct}). Thus, in $d$-dimensions, 
\bea
\fl
\la \d\rv^2 \ra = 2 d\left(D+\f{v_0^2}{(d-1)dD_r}\right)t - \f{v_0^2}{(d-1)^2 D_r^2}\left( 3-4\,e^{-(d-1)D_r t}+e^{-2(d-1)D_r t}\right)\nn\\
\fl 
+\f{2 D_v}{\gv(\gv+(d-1)D_r)} \left[ t -\f{1-e^{-(\gv+(d-1)D_r)t}}{(\gv+(d-1)D_r)} -\f{1-e^{-2\gv t}}{2\gv} + \f{e^{-2\gv t} - e^{-(\gv+(d-1)D_r) t}}{((d-1)D_r-\gv)}\right] . \nn\\
\label{eq:dr2_abp_speed_fluct}
\eea
\begin{figure*}[t]
\begin{center}
\includegraphics[width=8cm]{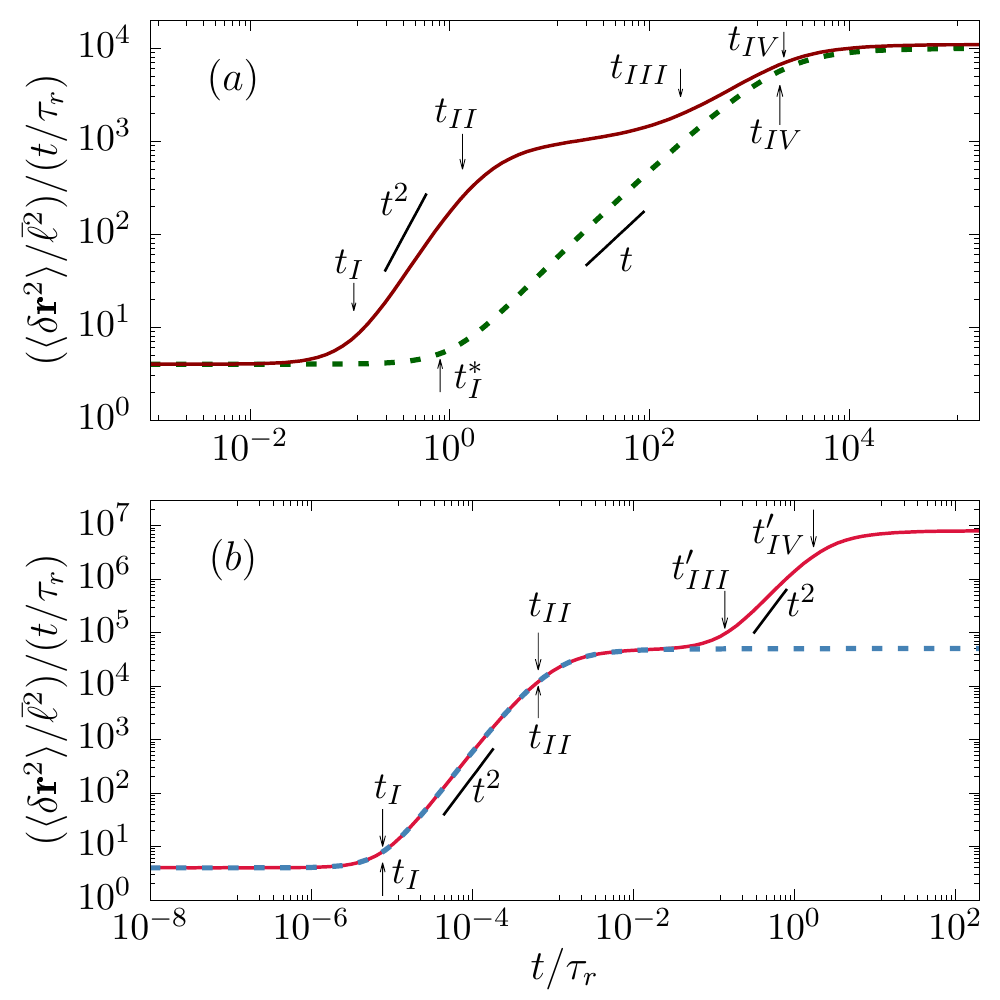} 
\caption{ (color online) Displacement fluctuations $\la \delta\rv^2\ra$ in equation~(\ref{eq:dr2_abp_speed_fluct}) as a function of time $t$ using $d=2$. $(a)$ Parameters used are $\tgv=5\times 10^{-4}$, $\tDv=2.5$ with active speed $Pe=22.36$ (solid line), $1.12$ (dashed line). The solid line shows four crossovers with crossover times $t_{I}/\t_r =0.008$, $t_{II}/\t_r =3.03$, $t_{III}/\t_r =202.8$ and $t_{IV}/\t_r =2.2\times 10^3$. The dashed line shows two crossovers with crossover times $t_{I}^{*}/\t_r =0.8$ and $t_{IV}/\t_r =2\times 10^3$. 
$(b)$ Parameters used are $\tgv=2\times 10^{3}$, $\tDv =10^{11}$ with $Pe=2\times 10^3$~(solid line), $10$ (dashed line). The solid line exhibits four crossovers with crossover times $t_{I}/\t_r=8\times 10^{-6}$, $t_{II}/\t_r=7\times 10^{-4}$, $t_{III}^{\prime}/\t_r=1.4\times 10^{-1}$ and $t_{IV}^{\prime}/\t_r=1.7$. The dashed line shows two crossovers with crossover times $t_{I}/\t_r=8\times 10^{-6}$, $t_{II}/\t_r=7\times 10^{-4}$. }  
\label{fig:dr2avg_abp_speed_fluct}
\end{center}
\end{figure*}
In the small time limit of $D_r t \ll 1$, $\gv t \ll 1$, 
expanding $\la\d\rv^2\ra$ in equation~(\ref{eq:dr2_abp_speed_fluct}) around $t=0$ leads to,
\bea
\fl
\la \d \rv^2 \ra = 2dDt+\f{2}{3}(D_v + (d-1)D_r v_0^2)t^3 -\f{1}{6} \left[ (d-1)D_r D_v + 3 D_v \g_v 
+ 3 (d-1)^2 D_r^2 v_0^2 \right]t^4
+{\cal O}(t^5). \nn\\
\label{eq:dr2_small_time_abp_speed_fluct}
\eea
The first two terms in the expansion shows a possible crossover from $\la\d\rv^2\ra\sim t$ to $\sim t^3$ at $t_{I}=[3dD/(D_v+(d-1)D_r v_0^2)]^{1/2}$. Moreover, a second crossover from $\la\d\rv^2\ra\sim t^3$ to $\sim t$ can appear at $t_{II} = 4(D_v +(d-1) D_r v_0^2)/((d-1)D_r D_v+3D_v \g_v+3v_0^2(d-1)^2D_r^2)$. 
In the long time  limit of $t\to\infty$, i.e., $D_r t \gg 1$, $\gv t \gg 1$, $\la\d\rv^2\ra$ in equation~(\ref{eq:dr2_abp_speed_fluct}) leads to a diffusive behavior
\bea
&\la \d\rv^2 \ra = \left(2dD+ \f{2 v_0^2}{(d-1)D_r}+ \f{2D_v}{\gv(\gv+(d-1)D_r)}\right) t .
\label{eq:dr2_long_time_abp_speed_fluct}
\eea
(i)~{\bf Slow  relaxation of active speed; $\g_v \ll (d-1)D_r$ }:
In the limit of $(d-1)D_r t\gg 1$ and $2\gv t\ll 1$, $\la\d\rv^2\ra$ in equation~(\ref{eq:dr2_abp_speed_fluct}) leads to
\bea
\fl
\la \d\rv^2 \ra = \left(2dD+ \f{2v_0^2}{(d-1)D_r}+ \f{4D_v}{((d-1)^2D_r^2 -\gv^2)}\right) t 
+ \f{2D_v}{((d-1)D_r - \gv)} t^2 + {\cal O}(t^3).
\label{eq:dr2_expansion_gv_lt_dr_abp_speed_fluct}
\eea
This allows a third crossover from $\la\d\rv^2\ra\sim t$ to $\la\d\rv^2\ra\sim t^2$ at $t_{III} \sim (2dD + 2v_0^2/(d-1)D_r +4D_v/((d-1)^2D_r^2-\gv^2))((d-1)D_r-\gv)/2D_v$. Finally, a crossover from $\la\d\rv^2\ra\sim t^2$ to $\sim t$ can appear at $t_{IV} \sim [2dD+ 2 v_0^2/(d-1)D_r + 2D_v/\{\gv(\gv+(d-1)D_r)\}]((d-1)D_r - \gv)/2D_v$. 

In the case of $t_{II}<t_{I}$, the number of possible crossovers reduces to two: from $\la\d\rv^2\ra\sim t$ to $\sim t^2$ to $\sim t$. Following a procedure similar to the analysis of crossovers in $\la\rv^2\ra$, we find that the first crossover from $\la\d\rv^2\ra\sim t$ to $\sim t^2$ appears at $t_{I}^{*}\sim 2dD((d-1)D_r-\gv)/2D_v$, obtained by comparing the first term in equation~(\ref{eq:dr2_small_time_abp_speed_fluct}) and the second term in equation~(\ref{eq:dr2_expansion_gv_lt_dr_abp_speed_fluct}). The second crossover $\la\d\rv^2\ra\sim t^2$ to $\sim t$ appears at $t_{IV}\sim [2dD+ 2 v_0^2/(d-1)D_r + 2D_v/\{\gv(\gv+(d-1)D_r)\}]((d-1)D_r - \gv)/2D_v$, obtained by comparing the second terms in equation~(\ref{eq:dr2_expansion_gv_lt_dr_abp_speed_fluct}) and equation~(\ref{eq:dr2_long_time_abp_speed_fluct}).

{In figure~\ref{fig:dr2avg_abp_speed_fluct}$(a)$ we show 
two examples of crossovers in $\la\d \rv^2\ra$ observed in 2d in the limit of $\gv\t_r\ll1$. We   identify the  crossover times in the figure. The figure is for parameter values $\tgv=\gv\t_r=5\times 10^{-4}$, $\tDv=D_v\t_r/\bar{v}^2=2.5$. 
The solid line, $Pe = v_0/\bar{v}= 22.36$ in figure~\ref{fig:dr2avg_abp_speed_fluct}$(a)$ exhibits all the four crossovers $\la\d\rv^2\ra\sim t$ to $\sim t^3$, to $\sim t$, to $\sim t^2$, to finally $\sim t$ as the requirement $t_{I}<t_{II}<t_{III}<t_{IV}$ is satisfied. The crossover times are $t_{I}/\t_r \sim [6/(Pe^2 +\tDv)]^{1/2}\approx 0.11$, $t_{II}/\t_r \sim 4(Pe^2 +\tDv)/(3Pe^2 + 3\tDv\tgv +\tDv)\approx 1.34$, $t_{III}/\t_r\sim [4+2Pe^2 +4\tDv/(1-\tgv^2)](1-\tgv)/2\tDv \approx 202.8$ and $t_{IV}/\t_r\sim [4+ 2 Pe^2 + 2\tDv/\{\tgv(1+\tgv)\}](1 - \tgv)/2\tDv\approx 2200$. 
The dashed line for $Pe = v_0/\bar{v}= 1.12$ in figure~\ref{fig:dr2avg_abp_speed_fluct}$(a)$ shows two crossovers $\la\d\rv^2\ra\sim t$ to $\sim t^2$ to $\sim t$ as $t_{II}<t_{I}$. The crossover times are $t_{I}^{*} /\t_r\sim 2(1-\tgv)/\tDv\approx 0.8$ and $t_{IV} /\t_r\approx 2\times 10^3$. Here, the first diffusive- ballistic crossover appears at $t_{I}^{*}$ as $t_{I}^{*}<t_{I}\approx 1.3 \, \t_r$.}

(ii)~{\bf Fast  relaxation of active speed; $\g_v \gg (d-1)D_r$ }:
In the other limit of $(d-1)D_r t\ll1$ and $2\gv t\gg1$, using $\exp({-2\gv t})= 0$, $\exp[\, -(\, (d-1)D_r+\gv)t\,]= 0$ and expanding $\exp(-D_r t)$ around $D_r t = 0$, equation~(\ref{eq:dr2_abp_speed_fluct}) leads to
\bea
\la \d\rv^2 \ra \simeq \left(2dD+\f{2D_v}{\g_v((d-1)D_r +\gv)}\right) t + \f{2}{3}(d-1) D_r v_0^2 t^3. 
\label{eq:dr2_expan_int_dr_lt_gv_abp_speed_fluct}
\eea
This predicts a third possible crossover from $\la\d\rv^2\ra\sim t$ to $\sim t^3$ at $t_{III} ^{\prime}\sim \left[3(dD+D_v/\{\gv((d-1)D_r+\gv)\})/(d-1)D_r v_0^2\right]^{1/2}$. The final crossover $\la\d\rv^2\ra\sim t^3$ to $\sim t$ can appear at $t_{IV}^{\prime} \sim \left[3(dD+v_0^2/(d-1)D_r+D_v/\{\gv((d-1)D_r+\gv)\})/(d-1)D_r v_0^2\right]^{1/2}$, with the crossover point obtained by comparing the second term in equation~(\ref{eq:dr2_expan_int_dr_lt_gv_abp_speed_fluct}) with equation~(\ref{eq:dr2_long_time_abp_speed_fluct}). If $t_{IV}^{\prime}\leq t_{III}^{\prime}$ these last two crossovers will not be possible. 

We demonstrate such crossovers in 2d, in the limit of $\gv\t_r\gg 1$, in figure~\ref{fig:dr2avg_abp_speed_fluct}$(b)$. The parameter values used are $\tgv=\gv\t_r=2\times 10^{3}$, $\tDv=D_v\t_r/\bar{v}^2=10^{11}$. 
The solid line in figure~\ref{fig:dr2avg_abp_speed_fluct}$(b)$ depicts the behavior at $Pe = v_0/\bar{v}=2\times 10^3$. This  exhibits all four crossovers from $\la\rv^2\ra\sim t$ to $\la\rv^2\ra\sim t^3$ to $\la\rv^2\ra\sim t$ to $\la\rv^2\ra\sim t^3$ to finally $\la\rv^2\ra\sim t$ as the requirement $t_{I}<t_{II}<t_{III}<t_{IV}$ is satisfied. 
In this case, the crossover points $t_{I}/\t_{r} \sim \left[6/(\tDv+Pe^2)\right]^{1/2}\approx 8\times 10^{-6}$, $ t_{II}/\t_r\sim 4(\tDv +Pe^2)/\left[\tDv+3\tDv \tgv +3 Pe^2\right]\approx 7\times 10^{-4}$, $ t_{III}^{\prime}/\t_r\sim \left[3[2+\tDv/\{\tgv(1+\tgv)\}]/Pe^2\right]^{1/2} \approx 1.4\times 10^{-1}$, and $ t_{IV}^{\prime}/\t_r \sim \left[3(2+Pe^2+\tDv/\{\tgv(1+\tgv)\})/Pe^2\right]^{1/2}\approx 1.7$ are identified in figure~\ref{fig:dr2avg_abp_speed_fluct}$(b)$. 
The dashed line corresponding to $Pe = v_0/\bar{v}= 10$ in figure~\ref{fig:dr2avg_abp_speed_fluct}$(b)$ shows two crossovers from $\la\d \rv^2\ra\sim t$, to $\la\d \rv^2\ra \sim t^3$ to finally $\la\d \rv^2\ra\sim t$.  As $t_{III}^{\prime} \approx t_{IV}^{\prime}\approx 27.4\,\t_r$, the corresponding crossovers from $\la\d\rv^2\ra\sim t$ to $\sim t^3$ to $\sim t$ is absent. The crossover times are $ t_{I}/\t_{r} \sim \left[6/(\tDv+Pe^2)\right]^{1/2}\approx 8\times 10^{-6}$ and $t_{II}/\t_r\sim 4(\tDv+Pe^2)/\left(\tDv+3\tDv\tgv +3 Pe^2\right)\approx 7\times 10^{-4}$.


\subsection{Components of displacement fluctuation}
\label{sec:components_disp_fluct_abp_speed_fluct}
The displacement of the ABP in parallel and  perpendicular directions with respect to the initial heading direction $\uv_0$ is studied here to identify any possible anisotropy in the dynamics.  The mean displacements are $\la \rpl \ra = \la \rv \ra \cdot \uv_0  \neq 0$ and $\la \rp \ra =  \la \rv \ra - \la \rpl \ra \uv_0$. Here $\la \rp \ra=0$ in the absence of external drive. In this Section, we compute the parallel and normal components of mean-squared displacemets and displacement fluctuations.

\subsubsection{Parallel component:}
We consider the initial active speed $v_1=v_0$.  Without any loss of generality, let us assume the initial heading direction of activity is towards the $x$-axis, $\uv_0 = \hat x$. 
We use equation~(\ref{eq:moment_recast}). 
Here $\psi=\rpl^2 = x^2$, giving $\la \psi\ra_0 = 0$, $\la\nabla_r^2 \psi\ra_s = 2\la 1\ra_s$, $\la\nabla_u^2 \psi \ra_s = 0$, $\la \p_{v}^{2}\psi\ra_s =0$, $\la (v-v_0)\p_{v} \psi\ra_s =0$, and $\la v \uv \cdot \nabla \psi\ra_s = 2\la v x u_x\ra_s$. Thus we find  
  \bea
   \la \rpl^2 \ra_s = \f{1}{s}\left[2D \la 1\ra_s+ 2\la v x u_x \ra_s \right] .\nn
  \eea 
To proceed we consider  $\psi = v x u_x$, giving $\la \psi\ra_0 = 0$, $\la\nabla_r^2 \psi\ra_s = 0$, $\la\nabla_u^2 \psi\ra_s = - (d-1) \la v x u_x\ra_s$, and $\la\uv \cdot \nabla \psi\ra_s = \la v u_x^2\ra_s$, 
leading to $\la v x u_x\ra_s = [\la v^2 u_x^2 \ra_s+\gv v_0 \la x u_x\ra_s]/(s  + (d-1) D_r + \gv)$. Further, 
$\la x u_x\ra_s = \la v u_x^2\ra_s/(s+(d-1)D_r)$. 
Further, we find, 
$\la v u_x^2\ra_s=\f{v_0 (s+2D_r)}{s(s+2dD_r)}$
and
\bea
\fl
\la v^2 u_x^2\ra_s=\f{v_0^2 (s+2D_r)}{s(s+2dD_r)} +\f{4 D_r D_v}{s(s+2\gv)(s+2dD_r)} 
+\f{2D_v (s+2\gv+2D_r)}{(s+2\gv)(s+2dD_r)(s+2\gv+2dD_r)}.\nn
\eea
Thus using these relations, we obtain 
\bea
\fl
\la \rpl^2 \ra_s = \f{2 D}{s^2} + \f{2 v_0^2 (s+2D_r)}{s^2 (s+(d-1)D_r)(s+2dD_r)} 
+\f{8 D_r D_v}{s^2(s+2\gv)(s+2dD_r)(s+\gv+(d-1)D_r)} \nn\\
\fl
+\f{4 D_v (s+2\gv+2D_r)}{s(s+2\gv)(s+2dD_r)(s+2\gv+2dD_r)(s+\gv+(d-1)D_r)}. 
\eea
The inverse Laplace transform gives,
\bea
\fl
\la \rpl^2 \ra = 2 \left(D + \f{v_0^2}{(d-1)dD_r}\right) t \nn\\
\fl
+ \f{v_0^2}{D_r^2}\left(\f{(d-1)e^{-2dD_r t}}{d^2 (d+1)}+\f{2(3-d)e^{-(d-1)D_r t}}{(d-1)^2 (d+1)}+\f{d^2-4d+1}{(d-1)^2 d^2}\right)\nn\\
\fl
+8D_r D_v \left[\f{-d^2 D_r^2 - 4 d\gv D_r + dD_r^2 -\gv^2 + \gv D_r}{8d^2 \gv^2 D_r^2 ((d-1)D_r + \gv)^2} +\f{t}{4d\gv D_r ((d-1)D_r + \gv)} \right]\nn\\
\fl
+\f{8D_r D_ve^{-2 d D_r t}}{8d^2 D_r^2 (dD_r -\gv)((d+1)D_r -\gv)}\nn\\
\fl
 -8D_r D_v  \f{e^{-((d-1)D_r +\gv)t}}{((d+1)D_r -\gv)((d-1)D_r -\gv)((d-1)D_r+\gv)^2}\nn\\
\fl
+8D_r D_v \f{e^{-2\gv t}}{8\gv^2 (dD_r -\gv)((d-1)D_r -\gv)} +4D_v \f{((d-1)D_r-\gv)e^{-2dD_r t}}{4d\gv D_r (dD_r-\gv)((d+1)D_r - \gv)}\nn\\
\fl
+4D_v\left[\f{D_r +\gv}{4d\gv D_r (dD_r+\gv)((d-1)D_r+\gv)}-\f{(d-1)e^{-(2dD_r+2\gv)t}}{4d\gv(dD_r+\gv)((d+1)D_r+\gv)} \right] \nn\\
\fl
+4D_v\left[\f{((3-d)D_r +\gv)e^{-((d-1)D_r +\gv)t}}{((d+1)^2 D_r^2 - \gv^2)((d-1)^2 D_r^2 -\gv^2)}-\f{e^{-2\gv t}}{4d\gv(dD_r-\gv)((d-1)D_r-\gv)}\right].\nn\\ 
\label{eq:r_para2_abp_speed_fluct}
\eea
Thus the parallel component of the displacement fluctuation $\la\d r_{\parallel}^2\ra =\la r_{\parallel}^{2}\ra-\la r_{\parallel}\ra^2$ is given by,
\bea
\fl
\la \delta\rpl^2 \ra = 2 \left(D + \f{v_0^2}{(d-1)dD_r}\right) t \nn\\
\fl
+ \f{v_0^2}{D_r^2}\left(\f{(d-1)e^{-2dD_r t}}{d^2 (d+1)}+\f{8 e^{-(d-1)D_r t}}{(d-1)^2 (d+1)}-\f{e^{-2(d-1)D_r t}}{(d-1)^2}-\f{4d-1}{(d-1)^2 d^2}\right)\nn\\
\fl
+8D_r D_v \left[\f{-d^2 D_r^2 - 4 d\gv D_r + dD_r^2 -\gv^2 + \gv D_r}{8d^2 \gv^2 D_r^2 ((d-1)D_r + \gv)} +\f{t}{4d\gv D_r ((d-1)D_r + \gv)} \right]\nn\\
\fl
+\f{8D_r D_ve^{-2 d D_r t}}{8d^2 D_r^2 (dD_r -\gv)((d+1)D_r -\gv)}\nn\\
\fl
-8D_r D_v  \f{e^{-((d-1)D_r +\gv)t}}{((d+1)D_r -\gv)((d-1)D_r -\gv)((d-1)D_r+\gv)^2}\nn\\
\fl
+8D_r D_v \f{e^{-2\gv t}}{8\gv^2 (dD_r -\gv)((d-1)D_r -\gv)} +4D_v \f{((d-1)D_r-\gv)e^{-2dD_r t}}{4d\gv D_r (dD_r-\gv)((d+1)D_r - \gv)}\nn\\
\fl
 +4D_v\left[\f{D_r +\gv}{4d\gv D_r (dD_r+\gv)((d-1)D_r+\gv)}-\f{(d-1)e^{-(2dD_r+2\gv)t}}{4d\gv(dD_r+\gv)((d+1)D_r+\gv)} \right] \nn\\
\fl
+4D_v\left[\f{((3-d)D_r +\gv)e^{-((d-1)D_r +\gv)t}}{((d+1)^2 D_r^2 - \gv^2)((d-1)^2 D_r^2 -\gv^2)}-\f{e^{-2\gv t}}{4d\gv(dD_r-\gv)((d-1)D_r-\gv)}\right].\nn\\
\label{eq:del_r_para2_abp_speed_fluct}
\eea

\begin{figure*}[t]
\begin{center}
\includegraphics[width=12cm]{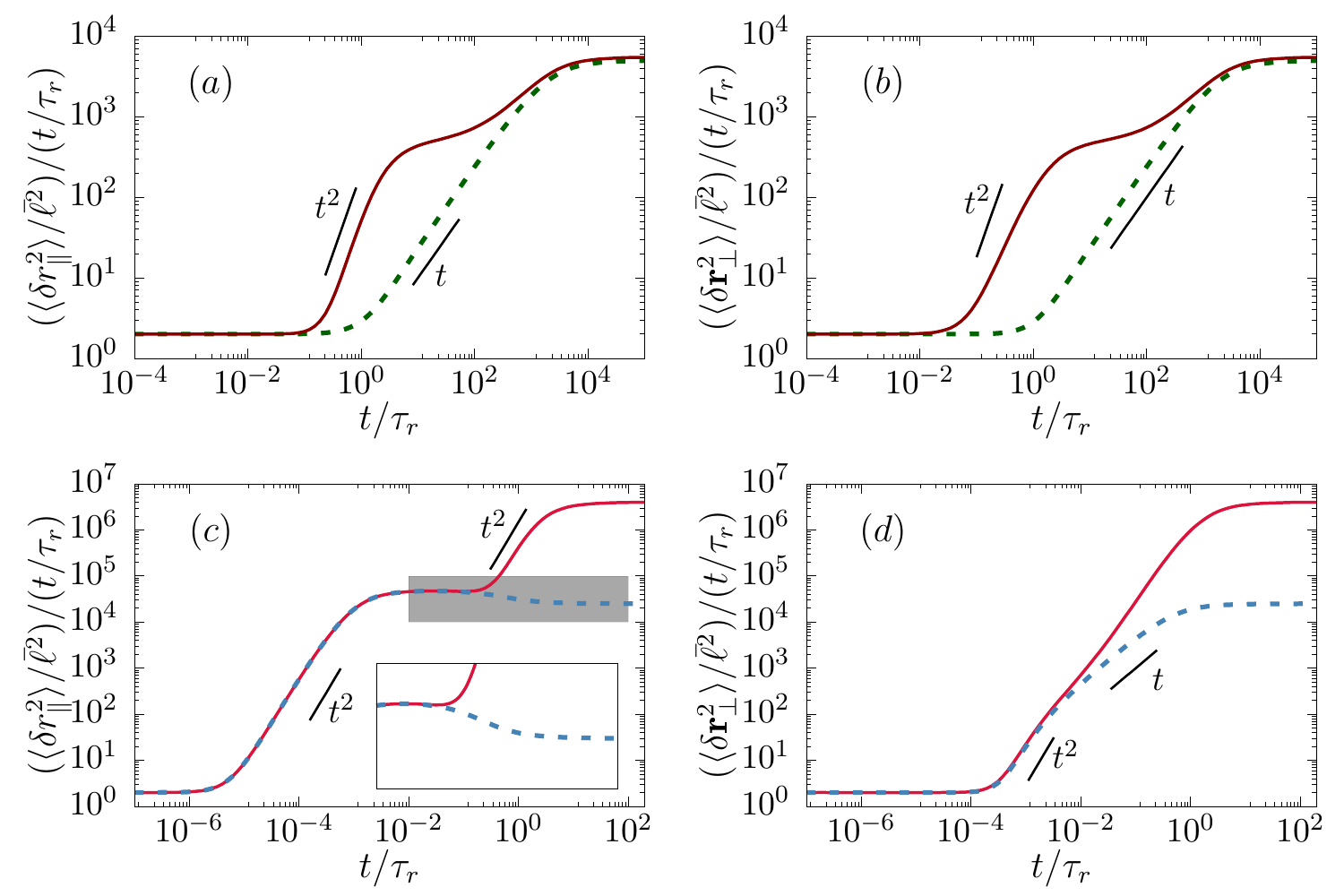} 
\caption{ (color online) Components of displacement fluctuation, $(a,c)$ $\la \d r_{\parallel}^2\ra$ and $(b,d)$ $\la \delta\rv_{\perp}^2\ra$  as a function of time $t$ in 2d. $(a,b)$ $\tgv=5\times 10^{-4}$, $\tDv=2.5$ with $Pe=22.36$ (solid line), $1.12$ (dashed line). $(c,d)$ $\tgv=2\times 10^{3}$, $\tDv=10^{11}$ with $Pe=2\times 10^3$ (solid line), $10$ (dashed line). The inset in Figure $(c)$ displaying a zoomed in view of the shaded region in the main figure  shows a sub-diffusive behavior in the parallel component of displacement fluctuation over an intermediate time regime. 
} 
\label{fig:dx2avg_y2avg_abp_speed_fluct}
\end{center}
\end{figure*}

\subsubsection{Perpendicular component:}
The fluctuations in the perpendicular component 
\bea
\fl
\la \d \rp^2 \ra =\la\d \rv^2\ra-\la\d r_{\parallel}^2\ra =  2 (d-1) \left(D + \f{v_0^2}{(d-1)dD_r}\right) t\nn\\
\fl
+ \f{v_0^2}{D_r^2}\left(\f{4 e^{-(d-1)D_r t}}{d^2 -1}-\f{(d-1) e^{-2 d D_r t}}{d^2 (d+1)} -\f{3d-1}{d^2 (d-1)}\right)\nn\\
\fl
+\f{2 D_v}{\gv(\gv+(d-1)D_r)} \left[ t -\f{1-e^{-(\gv+(d-1)D_r)t}}{(\gv+(d-1)D_r)} -\f{1-e^{-2\gv t}}{2\gv} + \f{e^{-2\gv t} - e^{-(d-1)D_r t}}{((d-1)D_r-\gv)}\right]\nn\\
\fl
-8D_r D_v \left[\f{-d^2 D_r^2 - 4 d\gv D_r + dD_r^2 -\gv^2 + \gv D_r}{8d^2 \gv^2 D_r^2 ((d-1)D_r + \gv)} +\f{t}{4d\gv D_r ((d-1)D_r + \gv)} \right]\nn\\
\fl
-\f{8D_r D_ve^{-2 d D_r t}}{8d^2 D_r^2 (dD_r -\gv)((d+1)D_r -\gv)}\nn\\
\fl
 +8D_r D_v \f{e^{-((d-1)D_r +\gv)t}}{((d+1)D_r -\gv)((d-1)D_r -\gv)((d-1)D_r+\gv)^2}\nn\\
\fl
-8D_r D_v\f{e^{-2\gv t}}{8\gv^2 (dD_r -\gv)((d-1)D_r -\gv)}-4D_v \f{((d-1)D_r-\gv)e^{-2dD_r t}}{4d\gv D_r (dD_r-\gv)((d+1)D_r - \gv)}\nn\\ 
\fl
-4D_v\left[\f{D_r +\gv}{4d\gv D_r (dD_r+\gv)((d-1)D_r+\gv)}-\f{(d-1)e^{-(2dD_r+2\gv)t}}{4d\gv(dD_r+\gv)((d+1)D_r+\gv)} \right] \nn\\
\fl
-4D_v\left[\f{((3-d)D_r +\gv)e^{-((d-1)D_r +\gv)t}}{((d+1)^2 D_r^2 - \gv^2)((d-1)^2 D_r^2 -\gv^2)}-\f{e^{-2\gv t}}{4d\gv(dD_r-\gv)((d-1)D_r-\gv)}\right] .\nn\\
\eea
In figure~\ref{fig:dx2avg_y2avg_abp_speed_fluct} we show various possible features of $\la \d r_{\parallel}^{2}\ra$ and $\la \d \rv^2_\perp\ra$ at different parameter regimes. ABPs without speed fluctuations display anisotropy in displacement fluctuations, with $\la \d r_{\parallel}^{2}\ra$ showing a crossover from $\sim t$ to $\sim t^4$, in contrast to the  $\sim t$ to $\sim t^3$ crossover found in  $\la \d \rv_{\perp}^{2}\ra$~\cite{Shee2020}. Similar asymmetry in the absence of thermal fluctuations was pointed out before in Ref.~\cite{Basu2018}.  As is shown in figure~\ref{fig:dx2avg_y2avg_abp_speed_fluct}($a$) and ($b$), such  a clear distinction in the components of displacement fluctuation can disappear in the presence of speed fluctuations. Depending on the $Pe$ value, both  $\la \d r_{\parallel}^{2}\ra$ and $\la \d \rv_{\perp}^{2}\ra$ can show $\sim t$ to $\sim t^2$ to $\sim t$ crossovers, or, $\sim t$ to $\sim t^3$ to $\sim t$ to $\sim t^2$ to $\sim t$ crossovers.  However, in the presence of large speed fluctuations and at larger $Pe$ the parallel component $\la \d r_{\parallel}^{2}\ra$ can even display sub-diffusive behavior in intermediate times~(figure~\ref{fig:dx2avg_y2avg_abp_speed_fluct}($c$)\,). This is qualitatively different from crossovers shown by $\la \d \rv_{\perp}^{2}\ra$ as is demonstrated in figure~\ref{fig:dx2avg_y2avg_abp_speed_fluct}($d$). 

The identification of multiple crossovers in the mean-squared displacement, displacement fluctuations, and  its components is the second main result of this paper.


\section{Fourth moment and kurtosis}
\label{sec:Fourth_moment_and_Kurtosis}
 In this Section, we present exact calculations for the fourth moments of active speed and displacement. The fourth moment of speed calculated from the Fokker-Planck equation is consistent with the underlying Gaussian process. The analytic predictions for the fourth moment of displacement shows agreement with direct numerical simulations. We further compute the kurtosis of displacement vector to capture the deviations from the Gaussian fluctuations. For these calculations we consider the initial active speed of the particle to be $v_1=Pe \bar v$ and the initial position at the origin. 
 
\subsection{Fourth moment of speed}
Using $\psi=v^4$ in equation~(\ref{eq:moment_recast}), we get, $\la v^4\ra_s =\left[v_0^4 + 12 D_v\la v^2\ra_s +4\gv v_0\la v^3\ra_s\right]/(s+4\gv)$ where, $\la v^2\ra_s=v_0^2/s + 2D_v /s(s+2\gv)$, $\la v^3\ra_s = v_0^3/s + 6D_v v_0/s(s+2\gv)$. This leads to,
\bea
&\la v^4\ra_s =\f{v_0^4}{s} + \f{12D_v v_0^2}{s(s+2\gv)} +\f{24D_v^2}{s(s+2\gv)(s+4\gv)}.\nn
\eea
Performing inverse Laplace transform we find 
\bea
&\la v^4\ra =v_0^4 + \f{6 D_v (\gv v_0^2 +D_v)}{\gv^2}\left(1-e^{-2\gv t}\right) - \f{3 D_v^2}{\gv^2}\left(1-e^{-4\gv t}\right).\nn
\eea
Writing $v=\d v+\la v\ra$,  Wick's theorem for a Gaussian process predicts $\la v^4\ra=\la v\ra^4 + 6\la v\ra^2 \la\d v^2\ra + 3\la\d v^2\ra^2$. The above expression agrees with this behavior. 

\subsection{Fourth moment of displacement}
Using $\psi = \rv^4$ in equation~(\ref{eq:moment_recast}), we get 
\bea
 &\la \rv^4 \ra_s = \f{1}{s}\left[4(d+2)D \la\rv^2\ra_s + 4 \la v(\uv \cdot \rv) \rv^2 \ra_s\right],
 \label{eq:r4ds_abp_speed_fluct}
 \eea
 where $\la\rv^2\ra_s$ has been already calculated in equation(\ref{eq:r2_Laplace_abp_speed_fluct}). Similarly, using equation~(\ref{eq:moment_recast}) we can calculate the various moments necessary to evaluate $\la \rv^4 \ra_s$. We list them below,
 \bea
 \fl
  \la v(\uv \cdot \rv)\rv^2\ra_s =\f{2(2+d)D\la v\uv \cdot \rv\ra_s+ \la v^2 \rv^2 \ra_s+ 2 \la v^2(\uv \cdot \rv)^2 \ra_s+\gv v_0\la (\uv\cdot\rv)\rv^2\ra_s}{s+(d-1)D_r+\gv},\nn
 \eea
 \bea
 \fl
\la v\uv\cdot\rv\ra_s\nn = \f{\la v^2\ra_s+\gv v_0 \la \uv\cdot\rv\ra_s}{s+(d-1)D_r+\gv}~,\nn\\
\fl
\la v^2\rv^2\ra_s = \f{2dD\la v^2\ra_s + 2 \la v^3 \uv\cdot\rv\ra_s +2 D_v \la \rv^2\ra_s +2\gv v_0 \la v \rv^2\ra_s}{s+2\gv},\nn\\
\fl
\la v^2 (\uv\cdot\rv)^2\ra_s = \f{2D\la v^2\ra_s + 2D_r \la v^2 \rv^2\ra_s + 2D_v \la (\uv\cdot\rv)^2\ra_s + 2\la v^3 \uv\cdot\rv\ra_s+ 2\gv v_0 \la v(\uv\cdot\rv)^2\ra_s}{s+2dD_r + 2\gv},\nn\\
\fl
\la (\uv\cdot\rv)r^2\ra_s = \f{2(2+d)D\la \uv\cdot\rv\ra_s + \la v \rv^2\ra_s +2\la v (\uv\cdot\rv)^2\ra_s}{s+(d-1)D_r},\nn
\eea
The calculations of these terms, in turn, require the following results,
\bea
\la v^2\ra_s=\f{v_0^2}{s} +\f{2D_v}{s(s+2\gv)}~,
~\la v^3\ra_s =\f{v_0^3}{s} + \f{6D_v v_0}{s(s+2\gv)}~,
\nn\\
\la\uv\cdot\rv\ra_s = \f{v_0}{s(s+(d-1)D_r)}~,~\la(\uv\cdot\rv)^2\ra_s =\f{2D\la 1\ra_s + 2D_r \la\rv^2\ra_s +2\la v \uv\cdot\rv\ra_s}{s+2dD_r},\nn\\
\la v^2 \uv\cdot\rv\ra_s = \f{2D_v\la \uv\cdot\rv\ra_s+\la v^3\ra_s + 2\gv v_0 \la v \uv\cdot\rv\ra_s}{s+(d-1)D_r+2\gv}~,\nn\\
\la v^3 \uv\cdot\rv\ra_s = \f{6D_v\la v (\uv\cdot\rv)\ra_s+\la v^4 \ra_s + 3\gv v_0 \la v^2 \uv\cdot\rv\ra_s}{s+(d-1)D_r+3\gv},\nn \\
\la v \rv^2\ra_s = \f{2dD\la v\ra_s + 2\la v^2 \uv\cdot\rv\ra_s + \gv v_0 \la\rv^2\ra_s}{s+\gv}~,\nn\\
\la v (\uv\cdot\rv)^2\ra_s = \f{2D\la v\ra_s + 2D_r \la v \rv^2\ra_s + 2 \la v^2 \uv\cdot\rv\ra_s +\gv v_0 \la(\uv\cdot\rv)^2\ra_s}{s+2dD_r+\gv},\nn
\eea

Finally, performing inverse Laplace transform of equation(\ref{eq:r4ds_abp_speed_fluct}) one obtains the expression for $\la\rv^4\ra(t)$. The expression is too lengthy to show here. Instead, we plot the expression for  $\la\rv^4\ra$ as a function of time in figure~\ref{fig:kurtosis_abp_speed_fluct}. In the following, we present the short and long time limit of $\la\rv^4\ra(t)$, and analyze its behavior.  
In the short time limit, an expansion of $\la\rv^4\ra$ around $t=0$ gives,
\bea
\fl
\la\rv^4\ra=4 d (d+2) D^2 t^2+4(d+2) D v_0^2 t^3 
+\left(v_0^4-\f{4(d+2)D}{3}\left((d-1)D_r v_0^2 - 2 D_v\right)\right)
t^4+\mathcal{O}(t^5). \nn\\
\eea
It shows that a $\la\rv^4\ra\sim t^2$ scaling at shortest time, which crosses over to $\la\rv^4\ra(t) \sim t^3$ scaling at $t_{I}=dD/v_0^2$. The crossover point is obtained by comparing the first two terms in the above expansion. A comparison between the second and third terms of the above expansion shows that a second crossover from $\la\rv^4\ra\sim t^3$ to $\sim t^4$ can appear at,
\bea
t_{II}=\f{12 (d+2) D v_0^2}{3 v_0^4 - 4(d+2) D ((d-1)D_r v_0^2 -2 D_v)},\nn
\eea 
provided $t_{II} > t_I$.  
In the long time limit, $\la\rv^4\ra$ approaches, 
\bea
\fl
\la\rv^4\ra\approx 
\f{4(2+d)\left[(d-1)D_r(D_v +dD\gv((d-1)D_r+\gv))+\gv((d-1)D_r+\gv)v_0^2\right]^2}{d (d-1)^2 D_r^2 \gv^2 ((d-1)D_r +\gv)^2}\, t^2. \nn
 \eea
In figure~\ref{fig:kurtosis_abp_speed_fluct}($a,b$), we show comparisons between the analytic expression for $\la\rv^4\ra$ and the direct numerical simulation result for this fourth moment to find clear agreement between them. 
Figure~\ref{fig:kurtosis_abp_speed_fluct}$(a)$ corresponds to the limit $D_r<<\gv$ and figure~\ref{fig:kurtosis_abp_speed_fluct}$(b)$ is plotted for parameter values obeying $D_r>>\gv$.
\subsection{Kurtosis: deviations from the Gaussian process}
\label{subsec:Kurtosis_abp_speed_fluct}
\begin{figure*}[t]
\begin{center}
\includegraphics[width=12cm]{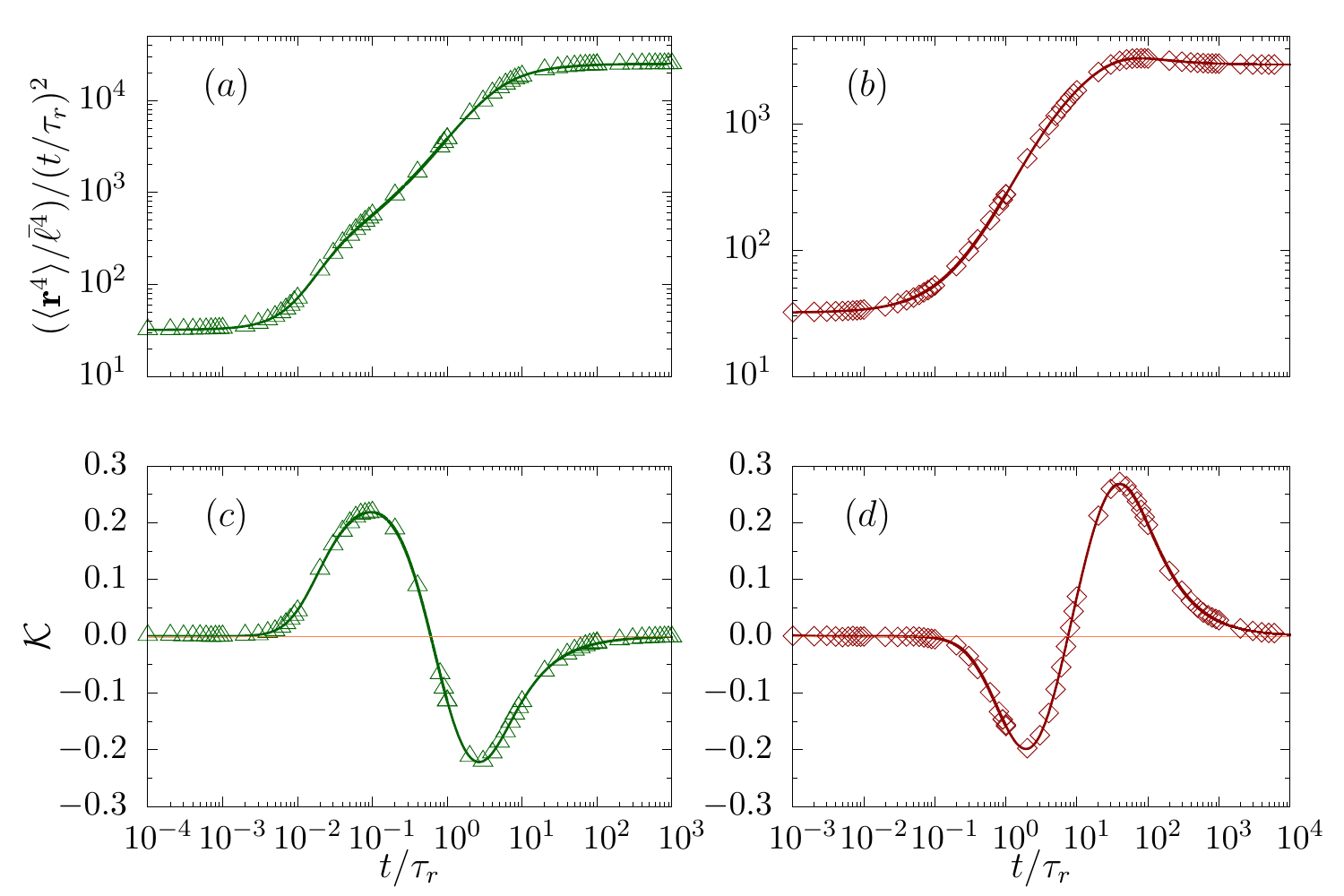} 
\caption{ (color online) Persistent motion: Plots of $\la\rv^4\ra$~($a,b$) and  Kurtosis ($\mathcal{K}$)~($c,d$) as a function of time in two dimensions. ($a,c$) Parameter values used are $\tgv=10^2$, $\tDv=4\times 10^4$, and $Pe =7.07$. 
($b,d$) Parameter values used are $\tgv =5\times 10^{-2}$, $\tDv=0.25$, and $Pe = 3.54$. 
The points denote simulation results averaged over $10^6$ independent trajectories. The solid lines depict analytic results obtained from the inverse Laplace transform of equation~(\ref{eq:r4ds_abp_speed_fluct}). The orange line in ($c,d$) corresponds to zero kurtosis. Initial conditions used are speed $v_1/\bar{v}=Pe$ and heading direction $\uv_0=\hat{x}$.} 
\label{fig:kurtosis_abp_speed_fluct}
\end{center}
\end{figure*}
For a Gaussian process with non-zero mean, the definition of the fourth moment of displacement can be expressed as,
\bea
\mu_4 : = \la\rv^2\ra^2 +\f{2}{d} \left(\la\rv^2\ra^2 -\la \rv\ra^4\right) .
\label{eq:mu4_gaussian}
\eea
Thus, deviations from such a Gaussian process is captured by the kurtosis 
\bea
\mathcal{K}=\f{\la\rv^4\ra}{\mu_4} -1 . 
\label{eq:kurtosis}
\eea
Figure~\ref{fig:kurtosis_abp_speed_fluct}$(c,d)$ shows the kurtosis as a function of time. A non-zero value of the kurtosis indicates deviations of the stochastic process from a possible Gaussian nature.    
A positive value appears for distributions with tails longer than normal distributions, while a negative value indicates a tail less extreme than the normal distributions.  
Figure~\ref{fig:kurtosis_abp_speed_fluct}$(c)$ corresponds to $\la\rv^4 \ra$ in figure~\ref{fig:kurtosis_abp_speed_fluct}$(a)$ in the limit of $D_r\gg\gv$. 
It shows deviations to positive values at shorter times and negative values at longer times before returning to the Gaussian behavior for long enough trajectories. The plot of kurtosis in  figure~\ref{fig:kurtosis_abp_speed_fluct}$(d)$ corresponds to $\la\rv^4 \ra$ in figure~\ref{fig:kurtosis_abp_speed_fluct}$(b)$  in the limit of $D_r\ll\gv$.
In contrast to figure~\ref{fig:kurtosis_abp_speed_fluct}($c$), in this parameter regime, the kurtosis shows deviations to negative values at shorter times that changes to positive values at longer times before returning to the Gaussian nature at the longest time scales. As it has been shown in Ref.~\cite{Shee2020}, the orientational fluctuations of the heading direction leads to  negative kurtosis in the intermediate times. The positive kurtosis observed here is determined by the active speed fluctuations that was not considered before.  In figure~\ref{fig:kurtosis_abp_speed_fluct}($c$) with $\tilde\gv \gg 1$, the orientational fluctuation time scale is longer than the speed fluctuation time scale. As a result, the negative kurtosis appears at a later time and the positive kurtosis at a shorter time. On the other hand, in figure~\ref{fig:kurtosis_abp_speed_fluct}($d$) with $\tilde\gv \ll 1$, the shorter orientational fluctuation time scale leads to the appearance of negative kurtosis at shorter times and positive kurtosis at longer times.  The presence of positive and negative kurtosis in the intermediate times due to the competition between orientational and speed relaxation is the third main result of this paper.

\section{Discussion} 
\label{sec:conclusions_ABP_speed_fluct}
In this paper we presented a detailed study of the dynamics of active Brownian particles with speed fluctuations, in the presence of thermal diffusion. In our model, two independent time scales describe the stochastic change of heading direction and speed. Here we considered the active speed generation using the Schienbein-Gruler model of simple energy pump~\cite{Schienbein1993, Schienbein1994}. We have extended the Fokker-Planck equation based method developed in Ref.~\cite{Hermans1952, Shee2020} to calculate all the relevant dynamical moments of motion in arbitrary dimensions. We presented some of these  calculations in detail.

First,  we calculated the mean-squared displacement in $d$-dimensions starting from the Fokker-Planck approach. The result we derived is consistent with the generic two-time auto-correlation function of active speed. 
In the limit of a fast relaxation, where the steady-state expression for the speed autocorrelation function can be used, our result for mean-squared displacement reduces to the earlier result obtained for 2d~\cite{Schienbein1994, Peruani2007}. Moreover, we calculated the fluctuations of the displacement vector, its components along and perpendicular to the initial  heading direction, and its fourth moment. These calculations showed several dynamical crossovers that we analyzed in detail and obtained expressions for the crossover times.  
The number of crossovers observed depends on the parameter values used. 
Finally, we calculated the kurtosis of the displacement vector to show deviations from the Gaussian process in intermediate times. The kurtosis deviates towards a positive value when the speed fluctuation dominates over the orientation fluctuation and a negative value when the orientation fluctuation dominates over the speed fluctuation. Thus, the kurtosis showed   opposite behaviors in the two limits of $\gv\t_r\ll1$ and $\gv\t_r\gg1$. Our predictions can be tested in experiments on artificial active particles, e.g., self-propulsion of Janus colloids~\cite{Bechinger2016, Howse2007, Jiang2010}. Our results can be useful in analyzing  the dynamics of motile cells having speed and directional fluctuations~\cite{Selmeczi2005, Frangipane2019, Otte2021}. 

\section*{Acknowledgments}
The numerical calculations were supported in part by SAMKHYA, the
high performance computing facility at Institute of Physics, Bhubaneswar. We thank Abhishek Dhar for useful discussions. D.C. thanks SERB, India for financial support through grant number MTR/2019/000750 and International Centre for Theoretical Sciences (ICTS) for an associateship. 

 \appendix
 \section{}
\section*{Steady state probability distribution of speed and its cumulative distribution}
\label{appendix:steady_state_prob_dist}

\begin{figure*}[t]
\begin{center}
\includegraphics[width=10cm]{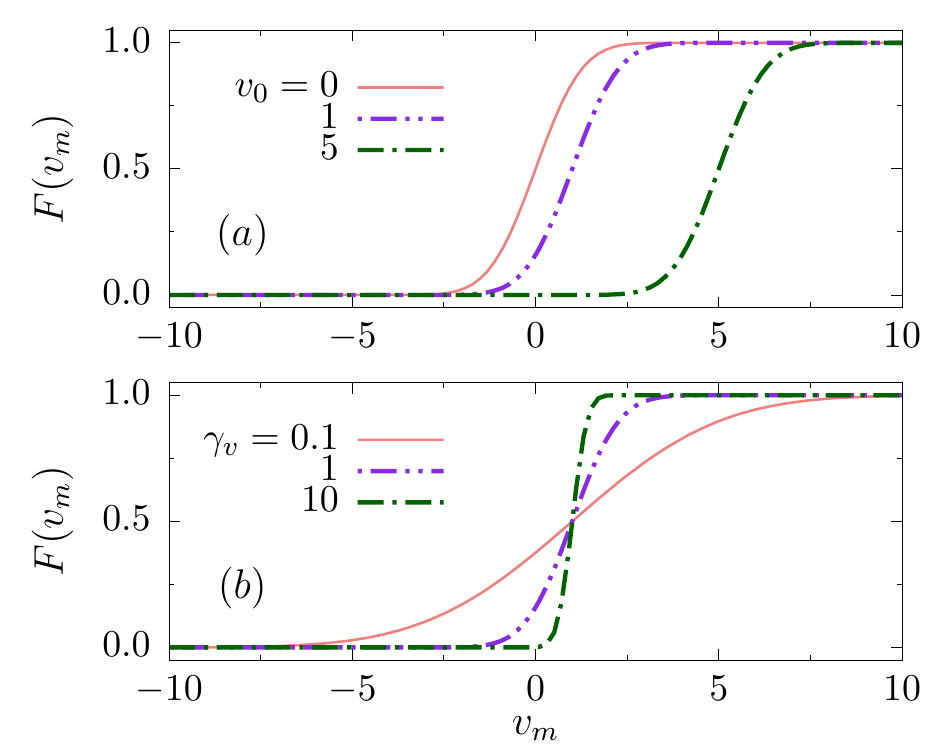} 
\caption{ (color online) Cumulative distribution function $F(v_m)$ in equation~(\ref{eq:cdf}) as a function of $v_m$ at $\tDv=1$. $(a)$ $Pe=0,1,5$ and $\tgv=1$. $(b)$ $\tgv=0.1,1,10$ and $Pe=1$.
} 
\label{fig:cdf}
\end{center}
\end{figure*} 

 The evolution equation of probability distribution of speed $P(v, t)$ derived from the Schienbein-Gruler mechanism~\cite{Schienbein1994} of active speed generation as in  equation~(\ref{eom:speed_abp_speed_fluct}), obeys the following Fokker-Planck equation 
 \bea
 \p_t P(v, t) = D_v \partial_{v}^{2} P + \gv \partial_v[(v-v_0)   P].
 \label{eq:Fokker-Planck_speed_evolution_equation}
 \eea
 The normalized steady-state distribution calculated from equation~(\ref{eq:Fokker-Planck_speed_evolution_equation}) has a Gaussian form peaked around the speed $v_0$,
 \bea
 P_{s}(v) = \left(\f{\gv}{2\pi D_v}\right)^{1/2} \exp\left(-\f{\gv}{2D_v} (v-v_0)^{2}\right)
 \label{eq:steady_state_speed_probability_distribution}
 \eea
The cumulative distribution function of speed up to a maximum value $v_m$ is 
\bea
F(v_m) &=& \left(\f{\gv}{2\pi D_v}\right)^{1/2} \int_{-\infty}^{v_m} dv~\exp\left(-\f{\gv}{2D_v} (v-v_0)^{2}\right) \nn\\
&=& \f{1}{2}\left[1+{\rm erf}\left(\f{v_m - v_0}{\sqrt{2D_v/\gv}}\right)\right] .
\label{eq:cdf}
\eea
In figure~\ref{fig:cdf}, we show the variation of cumulative distribution function for speed with changing $Pe=v_0/\bar v$ and $\tgv = \gv \t_r$. The probability of getting negative speed, an effective active speed opposite to the heading direction, decreases with increase of $v_0$ and $\gv$.

 \section{}
\section*{Autocorrelation of active speed}
\label{appendix:steady_state_corr_speed}
Here, we calculate the active speed auto-correlation function directly from the governing Langeving equation~(\ref{eom:speed_abp_speed_fluct}).
The formal solution of equation~(\ref{eom:speed_abp_speed_fluct}) with the initial condition $v(t=0)=v_1$ is 
\bea
v(t)=v_1\, e^{-\gv t} + \int_{0}^{t} \left(\gv v_0 + \sqrt{2D_v}\Lambda(t^{\prime})\right) e^{-\gv\left(t-t^{\prime}\right)}\, dt^{\prime},
\label{eq:formal_solution_speed}
\eea
with $\la \L (t) \ra=0$, and $\la \L (t) \L (t')\ra = \d(t-t')$. 
In this expression, the integration of the second term gives,
\bea
I= \int_{0}^{t} \gv v_0\, e^{-\gv\left(t-t^{\prime}\right)}\, dt^{\prime}= \gv v_0\,e^{-\gv t} \int_{0}^{t} e^{\gv t^{\prime}}\, dt^{\prime}=v_0 (1-e^{-\gv t}). \nn
\eea
This allows us to calculate the instantaneous mean speed 
\bea
\la v(t) \ra = v_1 e^{-\g_v t} + v_0 \left( 1 - e^{-\g_v t}\right). 
\eea
Thus, the deviation of speed from its mean value 
\bea
\d v(t) \equiv v(t)- \la v(t) \ra = \sqrt{2D_v} e^{-\gv t} \int_{0}^{t} \Lambda(t^{\prime}) e^{\gv t^{\prime}}\, dt^{\prime} .
\eea
As a result, the speed autocorrelation function of speed fluctuations can be calculated as
\bea
\la \d v(t_1) \d v(t_2)\ra = 2 D_v e^{-\gv\left(t_1 + t_2\right)} \int_{0}^{t_1} dt_1^{\prime} \int_{0}^{t_2}  dt_2^{\prime}\,  e^{\gv\left(t_1^{\prime} +t_2^{\prime}\right)} \d(t_1^{\prime}-t_2^{\prime}) 
\eea
If ($t_1 > t_2$), the $\d(t_1^{\prime}-t_2^{\prime})$ restricts the integration over $t_1^{\prime} = t_2^{\prime}$ line, then $t_1^{\prime}$ effectively runs up to $t_2$.
\bea
\la \d v(t_1) \d v(t_2)\ra = \f{D_v}{\gv} \left[e^{-\gv(t_1 -t_2)}-e^{-\gv(t_1+t_2)}\right]
\label{eq:autocorr_ABP_speed_fluct_A}
\eea

The steady state correlation may be obtained by, letting $t_1, t_2 \rightarrow \infty$ and keeping $\t=t_1 -t_2$ finite,
\bea
\la \d v(\tau) \d v(0)\ra = \f{D_v}{\gv} e^{-\gv\tau}.
\eea
In the steady state limit the instantaneous fluctuation, $\la \d v^2(0) \ra = D_v/\gv$.
Thus we may write, speed correlation in normalized form,
\bea
\f{\la \d v(\tau) \d v(0)\ra}{\la \d v^2(0)\ra} =  e^{-\gv\tau}
\label{eq:speed_corr_ABP_speed_fluct}
\eea
The fluctuation in speed can be derived from equation~(\ref{eq:autocorr_ABP_speed_fluct_A}) by setting $t_1 = t_2 \equiv t$,
\bea
\f{\la\d v^2 (t)\ra}{\la \d v^2(0)\ra}=\left(1-e^{-2\gv t}\right) .
\label{eq:speed_fluctuation_a}
\eea

\section*{References}

\bibliographystyle{unsrt}

\end{document}